\documentclass[english]{IEEEtran}
\usepackage[T1]{fontenc}
\usepackage[latin9]{inputenc}
\usepackage{color}
\usepackage{float}
\usepackage{amsmath}
\usepackage{amsthm}
\usepackage{amssymb}
\usepackage{stackrel}
\usepackage{graphicx}
\usepackage{setspace}

\makeatletter

\providecommand{\tabularnewline}{\\}
\floatstyle{ruled}
\newfloat{algorithm}{tbp}{loa}
\providecommand{\algorithmname}{Algorithm}
\floatname{algorithm}{\protect\algorithmname}

\theoremstyle{plain}
\newtheorem{thm}{\protect\theoremname}
\theoremstyle{remark}
\newtheorem{rem}[thm]{\protect\remarkname}

\usepackage[caption=false, font=footnotesize]{subfig}

\usepackage{algorithmic}
\usepackage{cite}

\makeatother

\usepackage{babel}
\providecommand{\remarkname}{Remark}
\providecommand{\theoremname}{Theorem}

\begin{document}
\title{Context-aware Constrained Reinforcement Learning Based Energy-Efficient
Power Scheduling for Non-stationary XR Data Traffic\vspace{-0.3cm}
}
\author{\singlespacing{}{\normalsize Kexuan Wang and An Liu, }{\normalsize\textit{Senior Member,
IEEE}}{\normalsize\thanks{Kexuan Wang and An Liu are with the College of Information Science
and Electronic Engineering, Zhejiang University, Hangzhou 310027,
China (email: \{kexuanWang, anliu\}@zju.edu.cn). \textit{(Corresponding
Author: An Liu)}}}\vspace{-0.8cm}
}
\maketitle
\begin{abstract}
In XR downlink transmission, energy-efficient power scheduling (EEPS)
is essential for conserving power resource while delivering large
data packets within hard-latency constraints. Traditional constrained
reinforcement learning (CRL) algorithms show promise in EEPS but still
struggle with non-convex stochastic constraints, non-stationary data
traffic, and sparse delayed packet dropout feedback (rewards) in XR.
To overcome these challenges, this paper models the EEPS in XR as
a dynamic parameter-constrained Markov decision process (DP-CMDP)
with a varying transition function linked to the non-stationary data
traffic and solves it by a proposed context-aware constrained reinforcement
learning (CACRL) algorithm, which consists of a context inference
(CI) module and a CRL module. The CI module trains an encoder and
multiple potential networks to characterize the current transition
function and reshape the packet dropout rewards according to the context,
transforming the original DP-CMDP into a general CMDP with immediate
dense rewards. The CRL module employs a policy network to make EEPS
decisions under this CMDP and optimizes the policy using a constrained
stochastic successive convex approximation (CSSCA) method, which is
better suited for non-convex stochastic constraints. Finally, theoretical
analyses provide deep insights into the CADAC algorithm, while extensive
simulations demonstrate that it outperforms advanced baselines in
both power conservation and satisfying packet dropout constraints.
\end{abstract}


\section{Introduction \label{sec:Introduction}}

Extended reality (XR) represents a category of emerging applications
in sixth-generation (6G) communications, characterized by orders of
magnitude larger packets than those in traditional low-latency applications
such as the Internet of Things (IoT) and Vehicle-to-Everything (V2X)
\cite{XR0large0packet}. Considering the latency-sensitive nature
of users, XR applications require each data packet must be successfully
transmitted within a \textit{hard}-latency constraint; otherwise,
the packet will be discarded \cite{XR}. Moreover, meeting this stringent
requirement often demands significant transmitting power resources,
which can lead to unaffordable economic costs and environmental crises
\cite{resource0consumption}. As a result, developing advanced energy-efficient
power scheduling (EEPS) algorithms to support XR has become a critical
research topic, attracting widespread attention from both academia
and industry \cite{EEbook,EE}.

Previous research efforts have extensively investigated low-latency
EEPS and other radio resource scheduling (RRS) \cite{SCAOPO,average-delay1,average-delay2},
however, most of them focused on reducing the average queue backlog
(\textit{soft}-latency constraint), which cannot avoid packet dropouts
in XR transmission. Only a few algorithms \cite{heuristic1-8,non-causal,AWGN,Poisson}
considered hard-latency constraints, and they still focused on designing
heuristics for a specific environment model. For example, \cite{heuristic1-8}
simply scheduled the packet closer to its deadline on a good channel
condition; a heuristic non-causal packet scheduler was used for the
causal case in \cite{non-causal}; analyses in \cite{AWGN} and \cite{Poisson}
were limited to AWGN channels and Poisson-distributed packet arrival
process, respectively. These traditional methods often lead to poor
performance and generalization ability in the real world. Recently,
\cite{model-free-CPO} modeled the RRS problem as a constrained Markov
Decision Process (CMDP) and solved it by a constrained deep Actor-Critic
(CDAC) algorithm, which belongs to a category of advanced constrained
reinforcement learning (CRL) algorithms. Without any prior information,
CDAC \cite{DRL,model-free-CPO,PPOLagTRPOLag} learns the environmental
model information in a model-free manner and optimize a policy for
making decisions online, achieving satisfactory simulated performance
in RRS and many other sequential decision-making problems, which greatly
inspires this paper. However, the existing CDAC algorithms are still
in need of further development to be applied to EEPS/other RRS tasks
in downlink XR transmission.

Modern CDAC algorithms usually employ deep neural networks (DNNs)
to avoid the curse of dimensionality in practical sequential decision
tasks with large continuous state and action space \cite{DRL}. This
approach introduces non-convexity into both the objective function
and the constraints of CRL problems \cite{SCAOPO}. Unfortunately,
most existing CDAC algorithms can only handle problems with simple
convex constraints. Specifically, there are two major categories of
policy optimization methods in existing CDAC algorithms \cite{SCAOPO}:
primal-dual methods \cite{PPOLagTRPOLag,NSGD1,primal_dual_conv2,ding2024last,SCAOPO18}
and CPO \cite{CPO,ProjectionCPO,firstorderCPO}. The primal-dual methods,
represented by trust region policy optimization-Lagrangian (TRPO-Lag)
and proximal policy optimization-Lagrangian (PPO-Lag) \cite{PPOLagTRPOLag},
casts the original problem into an unconstrained max-min saddle-point
problem and searches for optimal policies in the primal-dual domain.
Although \cite{reviewer1_1_2} proved that a general class of CRL
problems can be solved exactly in the convex dual domain, the method
of performing the non-convex optimization in the primal domain is
still open. Most existing primal-dual methods simply adopt the stochastic
gradient descent (SGD) method to update policy \cite{PPOLagTRPOLag,NSGD1,primal_dual_conv2},
which makes them only suitable for simple constraints where the feasible
set can be represented by a deterministic convex set. Although authors
in \cite{ding2024last} and \cite{SCAOPO18} applied optimistic gradient
descent methods to accelerate convergence, they still allowed the
agent to violate the safety constraints by oscillating around an optimal
safety policy. Authors in \cite{CPO,ProjectionCPO,firstorderCPO}
proposed the constrained policy optimization (CPO) algorithm, which
updates the policy within a trust-region framework. Nevertheless,
due to the non-convexity and high computation complexity, the CPO
introduces some approximate operations at the cost of sacrificing
the stable performance \cite{firstorderCPO} and also cannot assure
a feasible final result. In particular, our previous work \cite{SCAOPO}
considered non-convexity in the algorithm design, however, it simply
adopted the Monte Carlo (MC) method to estimate Q values and belongs
to the Actor-only algorithm, which suffers from a slow convergence.

Moreover, the non-stationary data traffic (packet arrival dynamics)
and sparse delayed feedback (rewards) for packet dropouts in downlink
XR transmission can also pose serious challenges to the existing CDAC
algorithms, which however, are generally ignored. Specifically, conventional
CDAC algorithms typically assume that the environment dynamics are
stationary \cite{meta0learning1}. However, in XR scenarios, the packet
arrival probabilities and distributions of packet lengths continuously
change due to their close relationship with the type of each user's
online activities and the peak and off-peak hours \cite{nonstationaryURLLC}.
As a result, conventional CDAC algorithms must learn the environmental
model information from scratch whenever the packet arrival dynamics
undergo changes, which may require a long adaptation time. Meta-learning
is a category of methods that can enhance the adaptability of machine
learning (ML) algorithms in various tasks/dynamic environments, divided
into gradient-based meta-learning and context-based meta-learning
\cite{MAML}. The gradient-based methods \cite{MAML,MAML-CPO,gradientbased}
attempt to learn effective initial parameters by exploring the common
structures of various tasks/environment dynamics, thereby reducing
the convergence time of the algorithms. The context-aware methods
\cite{PERAL,context3,VAE} encode the features of each task/environment
dynamic into a specific latent variable to guide ML algorithms towards
faster convergence. Because hidden variables can represent both similarities
and differences between different environmental dynamics, context-aware
methods tend to be better suited to a rapidly changing environment
than gradient-based methods \cite{context3}. An initial exploration
to bridge a typical gradient-based meta-learning method, i.e., MAML,
into CPO is made in \cite{MAML-CPO}. However, to the best of our
knowledge, the context-aware meta-learning enabled CDAC in a non-stationary
environment has not been investigated. In addition, because the sparse
rewards for packet dropouts can only be received when the hard-latency
constraints are reached, the existing methods all tend to be inefficient
or fail in XR transmission, which makes the problem even trickier.

To combat these challenges above, this paper brings up a context-aware
constrained reinforcement learning (CACRL) algorithm to support the
energy-efficient power scheduling (EEPS) task in XR downlink transmission
with non-stationary XR data traffic. Note that this algorithm is also
easy to extend to scheduling other radio resources with trivial modifications.
The main contributions are as follows:
\begin{itemize}
\item \textbf{More realistic and tractable problem formulation:} We formulate
the EEPS task in XR downlink transmission as a dynamic parameter-constrained
Markov Decision Process (DP-CMDP). By using long-term power consumption
and packet dropout rates as performance metrics, the DP-CMDP formulates
the requirements of XR applications appropriately. Moreover, by adopting
a transition distribution defined by dynamic parameters, the DP-CMDP
can model the non-stationary nature of XR data traffic, a feature
overlooked in the genreal CMDP.
\item \textbf{An advanced CACRL Algorithm:} We propose an advanced CACRL
algorithm consisting of a context inference (CI) module and a CRL
module to address the challenging DP-CMDP. Specifically, by integrating
a context-aware meta-learning method and a potential-based reward
reshaping method, the CI module utilizes an encoder to characterize
the current transition function according to recent observation samples
(which is the so-called context), and employs potential networks to
transform the original packet dropout rewards into dense immediate
ones. These operations transform the original DP-CMDP problem into
a general CMDP problem. The CRL module then employs a stochastic policy
network to make EEPS decisions under this CMDP and, to better satisfy
non-convex stochastic constraints, optimizes the policy using a constrained
stochastic successive convex approximation (CSSCA) method, whose main
idea is to transform the original non-convex stochastic constraint
problem into a series of convex surrogate objectives and feasible
optimization problems. Note that the networks in both modules are
jointly optimized in model-free manners, without requiring any prior
knowledge of the environment model. Special designs are also adopted
to minimize the training overhead as much as possible.
\item \textbf{Theoretical and simulation analyses:} Comprehensive theoretical
analyses are provided to offer deep insights into the convergence
of the proposed CACRL algorithm. Moreover, sufficient ablation and
comparative simulation results further validate the effectiveness
of the CI module and the CRL module, demonstrating that the proposed
algorithm outperforms advanced baselines both in power conservation
and meeting packet dropout constraints.
\end{itemize}

\section{System Model and Problem Formulation}

\subsection{System Model}

\begin{figure}
\begin{centering}
\includegraphics[width=7.8cm,height=4cm]{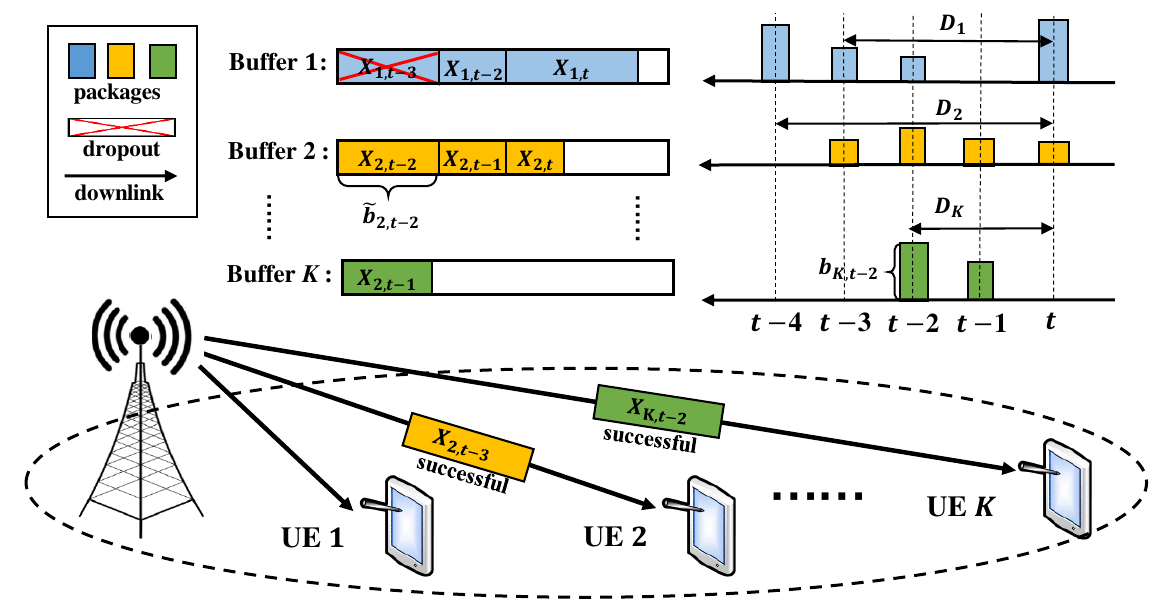}\vspace{-0.3cm}
\caption{\label{fig:System-Model}System model and timeslot diagram}
\par\end{centering}
\vspace{-0.6cm}
\end{figure}

We consider downlink transmission from an $M$-antennas base station
(BS) to $K$ single-antenna XR users, as illustrated in Fig. \ref{fig:System-Model}.
At the BS, each user $k$ is assigned a separate buffer $k$. Data
packets from some higher-layer applications randomly arrive at the
buffers and are stored in the form of queues. At the same time, the
BS dynamically schedules the transmission power resource for the downlink
transmission of each data queue. In the following, we provide further
details of the data queue model and the downlink transmission process.

\subsubsection{Data Queues Model}

In this subsection, we describe the dynamic characteristics of the
data queues. We divide time into some $\tau_{0}$s-timeslots and assume
that the packet $X_{k,t}$ of length $b_{k,t}$ for the $k$-th user
arrives from upper-layer applications to the BS with a probability
$P_{k}\in\bigl(0,1\bigr)$ at the $t$-th timeslot. Note $P_{k,t}$
and the distribution of $b_{k,t}$ are typically unknown to the BS.
Moreover, since the packet arrival processes are closely related to
the frequently switching active users and their ongoing online activities,
we assume that potential changes may occur in $P_{k,t}$ and the distribution
of $b_{k,t}$ at each timeslot $t$. The frequency of changes depends
on the specific scenario.

The arrived data packets are served according to a first-come-first-served
protocol. Considering the latency-sensitive nature of XR users, each
packet $X_{k,t}$ is assigned a \textit{hard}-latency constraint $D_{k}$,
i.e., all data packets $\bigl\{ X_{k,t-t_{d}}\bigr\}_{t_{d}=D_{k},D_{k}+1,\ldots}$
have been successfully transmitted or dropout. Denoting the remaining
length of packet $X_{k,t}$ as $\bar{b}_{k,t}$, we use a binary function
$\mathbb{I}_{k,t}$ to indicate whether a packet dropout occurs at
the buffer $k$ at timeslot $t$:
\[
\mathbb{I}_{k,t}^{\mathrm{drop}}\triangleq\begin{cases}
1, & \mathrm{if}\ \bar{b}_{k,t-D_{k}}>\tau_{0}\mathcal{R}_{k,t-1},\\
0, & \mathrm{otherwise},
\end{cases},\forall k,t
\]
where $\mathcal{R}_{k,t}$ denotes the transmission rate assigned
to buffer $k$ at timeslot $t$. Both transmission power shortage
and bad channel conditions can lead to the occurrences of packet dropout.
For ease of understanding, we show a possible state of the data queues
in Fig. \ref{fig:System-Model}.

\subsubsection{Downlink Transmission Process}

At each timeslot $t$, the downlink channel state information (CSI)
matrix is denoted as $\mathbf{H}_{t}\triangleq\bigl[\boldsymbol{h}_{1,t};\boldsymbol{h}_{2,t};\ldots;\boldsymbol{h}_{K,t}\bigr]\in\mathbb{C}^{K\times M_{t}}$,
where $\boldsymbol{h}_{k,t}\in\mathbb{R}^{1\times M}$ represents
the channel from the BS to the $k$-th user. It is assumed that $\mathbf{H}_{t}$
is block flat fading. For convenience, we also denote a queue state
information (QSI) matrix $\mathbf{B}_{t}\in\mathbb{R}^{2\times\sum_{k=1}^{K}D_{k}}\triangleq\bigl[\boldsymbol{b}_{1,t},\ldots,\boldsymbol{b}_{K,t};\boldsymbol{\bar{b}}_{1,t},\ldots,\bar{\boldsymbol{b}}_{K,t}\bigr]$,
where vectors $\boldsymbol{b}_{k,t}\in\mathbb{R}^{1\times D_{k}}\triangleq\bigl[b_{k,t},\ldots,b_{k,t-D_{k}+1}\bigr]$
and $\boldsymbol{\bar{b}}_{k,t}\in\mathbb{R}^{1\times D_{k}}\triangleq\bigl[\bar{b}_{k,t},\ldots,\bar{b}_{k,t-D_{k}+1}\bigr]$
record the original lengths and remaining lengths of packets $\bigl\{ X_{k,t-t_{d}}\bigr\}_{t_{d}=0,1,\ldots,D_{k}-1}$,
respectively.

According to the current CSI $\mathbf{H}_{t}$ and QSI $\mathbf{B}_{t}$,
the BS schedules power $p_{k,t}$ for user $k,\forall k$ for downlink
transmission, where we denote $\boldsymbol{p}_{t}\triangleq\bigl[p_{1,t},p_{2,t},\ldots,p_{K,t}\bigr]\in\mathbb{R}^{1\times K}$.
Then, the transmit signal $\boldsymbol{x}_{t}\in\mathbb{C}^{K\times1}\triangleq\bigl[x_{1,t},x_{2,t},\ldots,x_{K,t}\bigr]^{\intercal}$
satisfying $\mathbb{E}\bigl[\boldsymbol{x}_{t}\boldsymbol{x}_{t}^{\intercal}\bigr]=\mathrm{Diag}\bigl(\boldsymbol{p}_{t}\bigr)$
is processed by precoding matrix $\mathbf{V}_{t}\in\mathbb{C}^{M_{t}\times L}\triangleq\bigl[\boldsymbol{v}_{1,t},\boldsymbol{v}_{2,t},\ldots,\boldsymbol{v}_{K,t}\bigr]$.
Further, we calculate $\mathbf{V}_{t}$ by the normalized regularized
zero-forcing (RZF):
\begin{align*}
\mathbf{V}_{t} & =\mathbf{H}_{t}^{H}\bigl(\mathbf{H}_{t}\mathbf{H}_{t}^{H}+\epsilon_{t}\mathbf{I}\bigr)^{-1}\mathbf{\Lambda}^{1/2},\text{\ensuremath{\forall t}},
\end{align*}
where $\mathbf{\Lambda}^{1/2}=\mathrm{Diag}\bigl(\bigl[\left\Vert \boldsymbol{\bar{v}}_{1,t}\right\Vert ^{-1},\left\Vert \boldsymbol{\bar{v}}_{2,t}\right\Vert ^{-1},\ldots,\left\Vert \boldsymbol{\bar{v}}_{K,t}\right\Vert ^{-1}\bigr]\bigr)$
is the normalization matrix, $\boldsymbol{\bar{v}}_{k,t}$ is the
$l$-th column of $\bar{\mathbf{V}}_{t}\triangleq\mathbf{H}_{t}^{H}\bigl(\mathbf{H}_{t}\mathbf{H}_{t}^{H}+\epsilon_{t}\mathbf{I}\bigr)^{-1}$,
and $\epsilon_{t}$ is the regularization factor. Note that the RZF
precoder is widely used in practical systems and is asymptotically
optimal for large $M_{t}$ and/or high SNR \cite{RZF}. Further, the
received signal $y_{k,t}\in\mathbb{C}$ is given by
\[
y_{k,t}=\boldsymbol{h}_{k,t}\boldsymbol{v}_{k,t}x_{k,t}+\sum_{m\neq k}^{K}\boldsymbol{h}_{k,t}\boldsymbol{v}_{m,t}x_{m,t}+n_{l},\forall t,l,
\]
where $n_{k}\sim\mathcal{CN}\bigl(0,\sigma_{k}^{2}\bigr)$ represents
the additive white Gaussian noise. Finally, the transmission rate
$\mathcal{R}_{k,t}$ is given by
\[
\mathcal{R}_{k,t}=W\mathrm{log}_{2}\Bigl(1+\frac{p_{k,t}\bigl|\bigl(\boldsymbol{h}_{k,t}\bigr)^{H}\boldsymbol{v}_{k,t}\bigr|^{2}}{\sum_{m\neq k}p_{m,t}\bigl|\bigl(\boldsymbol{h}_{k,t}\bigr)^{H}\boldsymbol{v}_{m,t}\bigr|^{2}+\sigma_{k}^{2}}\Bigr),\forall k,t
\]
where $W$ denotes the bandwidth.

\subsection{Problem Formulation}

Considering the characteristics and requirements of XR applications,
we model the EEPS task as a dynamic parameter-constrained Markov decision
process (DP-CMDP), which can be denoted as a sequence of tuples $\bigl\{\mathcal{M}^{\boldsymbol{z}_{t}}=\left(\mathcal{S},\mathcal{\mathcal{A}},P_{\mathrm{T}}^{\boldsymbol{z}_{t}},R,C\right)\bigr\}_{t=0,1,\ldots}$
with the following definitions:
\begin{itemize}
\item \textbf{The state space} $\mathcal{S}$\textbf{: }the QSI space and
CSI space constitute the state space $\mathcal{S}$, i.e., $\boldsymbol{s}_{t}\triangleq\bigl\{\mathrm{vec}\bigl(\mathbf{B}_{t}\bigr)^{T},\mathrm{vec}\bigl(\mathbf{H}_{t}\bigr)^{T}\bigr\}$.
\item \textbf{The action space} $\mathcal{A}$\textbf{: }the action consists
of the scheduling decision and the regularization factor, i.e, $\boldsymbol{a}_{t}\triangleq\bigl\{\boldsymbol{p}_{t},\epsilon_{t}\bigr\}\in\mathcal{A}$.
Specifically, $\boldsymbol{a}_{t}$ is sampled according to a stochastic
policy $\pi$. To overcome the curse of dimension, the policy $\pi$
is parameterized by network $\pi_{\boldsymbol{\theta}}$ with $n_{\boldsymbol{\theta}}$-dimensional
parameter $\boldsymbol{\theta}\in\Theta$.
\item \textbf{The dynamic transition function $P_{\mathrm{T}}^{\boldsymbol{z}_{t}}$:}
at each timeslot $t$, the transition function $P_{\mathrm{T}}^{\boldsymbol{z}_{t}}$
is defined by a latent variable $\boldsymbol{z}_{t}$, where $\boldsymbol{z}_{t}\in\mathcal{Z}$
is unobservable and determined by the current underlying packet arrival
dynamics. The value $P_{\mathrm{T}}^{\boldsymbol{z}_{t}}\left(\boldsymbol{s}_{t+1}\mid\boldsymbol{s}_{t},\boldsymbol{a}_{t}\right)$
denotes the probability of transition from $\boldsymbol{s}_{t}$ to
$\boldsymbol{s}_{t+1}$ upon taking $\boldsymbol{a}_{t}$. Considering
the non-stationary nature of XR data traffic, we assume that $\boldsymbol{z}_{t}$
shifts stochastically according to an unknown distribution $P_{\mathrm{shift}}\bigl(\boldsymbol{z}_{t}\bigr|\boldsymbol{z}_{t-1}\bigr)$,
where $\boldsymbol{z}_{t}\neq\boldsymbol{z}_{t-1}$ indicates that
packet arrival dynamics have changed. Consequently, the sequence $\bigl\{\boldsymbol{z}_{t}\bigr\}_{t=0,1,\ldots}$is
a Markov random process.
\item \textbf{The reward }$R$\textbf{ and }$C$: we define reward functions
$R\left(\boldsymbol{s}_{t},\boldsymbol{a}_{t}\right)\triangleq\stackrel[k=1]{K}{\sum}p_{k,t}$
and $C_{k}\left(\boldsymbol{s}_{t},\boldsymbol{a}_{t}\right)\triangleq\mathbb{I}_{k,t}^{\mathrm{drop}}$,
which respectively represent the transmission power consumption and
the number of packet dropouts of each user at the $t$-th timeslot.
Please note that since the occurrences of packet dropouts are only
known when the hard-latency constraints are reached, rewards\textbf{
}$C$ are \textsl{sparse} and \textsl{delayed}, i.e., $\bigl\{ C_{k}\bigl(\boldsymbol{s}_{t},\boldsymbol{a}_{t}\bigr)\bigr\}_{t=0,1,\ldots},$$\forall k$,
is a 01 sequence with relatively fewer occurrences of 1.
\end{itemize}
Note that if $\boldsymbol{z}_{t}$ is known at each timeslot $t$,
the original DP-CMDP can be transformed into a stationary CMDP $\mathcal{M}\triangleq\bigl(\dot{\mathcal{S}},\mathcal{\mathcal{A}},P_{\mathrm{T}},R,C\bigr)$
by defining an augmented state space $\dot{\mathcal{S}}\triangleq\mathcal{S}\times\mathcal{Z}$,
where the transition function $P_{\mathrm{T}}$ remain unchanged,
with the probability $P_{\mathrm{T}}\bigl(\dot{\boldsymbol{s}}_{t+1}\mid\dot{\boldsymbol{s}}_{t},\boldsymbol{a}_{t}\bigr)=P_{\mathrm{T}}^{\boldsymbol{z}_{t}}\left(\boldsymbol{s}_{t+1}\mid\boldsymbol{s}_{t},\boldsymbol{a}_{t}\right)P_{\mathrm{shift}}\bigl(\boldsymbol{z}_{t+1}\bigr|\boldsymbol{z}_{t}\bigr)$.

This paper aims to design an algorithm that can, on the one hand,
continuously infer $\bigl\{\boldsymbol{z}_{t}\bigr\}_{t=0,1,2,\ldots}$,
and on the other hand, obtain a policy $\pi_{\boldsymbol{\theta}}$,
where $\pi_{\boldsymbol{\theta}}\bigl(\boldsymbol{a}_{t}\mid\dot{\boldsymbol{s}}_{t}\bigr)$
represents the probability of sampling the action $\boldsymbol{a}_{t}$
given the augmented state $\dot{\boldsymbol{s}}_{t}=\bigl\{\boldsymbol{s}_{t},\boldsymbol{z}_{t}\bigr\}$,
by solving the following non-convex stochastic optimization problem
under the transformed CMDP:
\begin{align}
\underset{\theta\in\Theta}{\mathrm{min}}f_{0}\left(\boldsymbol{\theta}\right)\overset{\triangle}{=} & \lim_{T\rightarrow\infty}\mathbb{E}\Bigl[\sum_{t=0}^{T-1}C_{0}^{'}\left(\boldsymbol{s}_{t},\boldsymbol{a}_{t}\right)\Bigr]\label{eq:CMDPs}\\
\mathrm{s.t.}f_{k}\left(\boldsymbol{\theta}\right)\overset{\triangle}{=} & \lim_{T\rightarrow\infty}\mathbb{E}\Bigl[\frac{1}{T}\sum_{t=0}^{T-1}C_{k}^{'}\left(\boldsymbol{s}_{t},\boldsymbol{a}_{t}\right)\Bigr]\leq0,\forall k,\nonumber 
\end{align}
where we denote $C_{0}^{'}\left(\boldsymbol{s}_{t},\boldsymbol{a}_{t}\right)=-R\left(\boldsymbol{s}_{t},\boldsymbol{a}_{t}\right)$
and $C_{k}^{'}\left(\boldsymbol{s}_{t},\boldsymbol{a}_{t}\right)=C_{k}\left(\boldsymbol{s}_{t},\boldsymbol{a}_{t}\right)-c_{k}$
with $c_{k}$ representing the allowed maximum packet dropout rate
for user $k$. The notation $\mathbb{E}\bigl[\cdot\bigr]$ denotes
the expectation taken over $\dot{\boldsymbol{s}}_{t+1}\sim P_{\mathrm{T}}\bigl(\dot{\boldsymbol{s}}_{t+1}\mid\dot{\boldsymbol{s}}_{t},\boldsymbol{a}_{t}\bigr)$
and $\boldsymbol{a}_{t}\sim\pi_{\boldsymbol{\theta}}\left(\boldsymbol{a}_{t}\mid\dot{\boldsymbol{s}}_{t+1}\right)$.
The goal of problem (\ref{eq:CMDPs}) is to minimize long-term average
power consumption while ensuring that the packet dropout rate for
each user $k$ does not exceed $c_{k}$.
\begin{rem}
However, designing such an algorithm is non-trivial. The main reasons
are as follows: 1) most existing algorithms are designed for a stationary
CMDP without considering the non-stationary nature of the real-world
scenarios. Therefore, methods that can effectively infer the latent
variables are still lacking, especially in XR scenarios with sparse
delayed feedback rewards $\bigl\{ C_{k}\bigl(\boldsymbol{s}_{t},\boldsymbol{a}_{t}\bigr)\bigr\}_{t=1,2,\ldots},$$\forall k$;
2) even if the latent variables can be obtained, it is quite challenging
for an algorithm to ensure that the non-convex stochastic constraints
are satisfied at a limiting point.
\end{rem}

\section{The Context-aware Constrained Reinforcement Learning Algorithm}

\subsection{Outline of the CACRL Algorithm}

\begin{figure}
\centering{}\includegraphics[width=8cm,height=4.6cm]{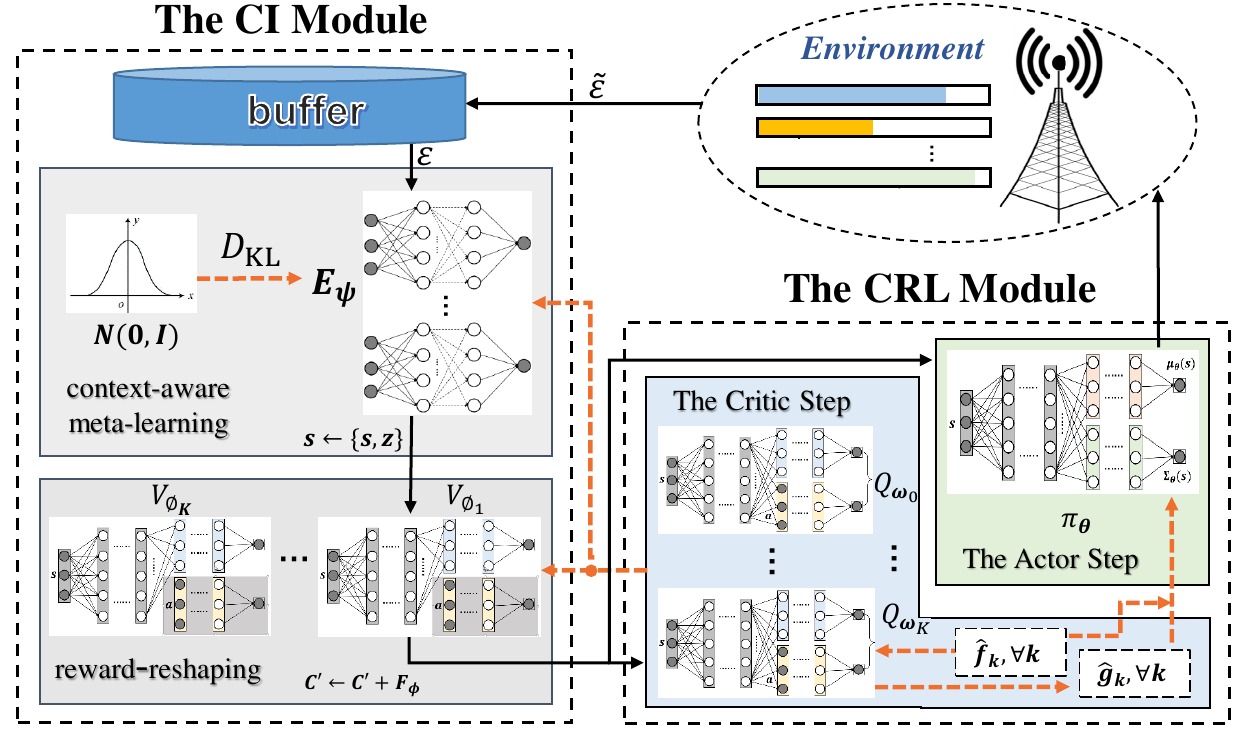}\vspace{-0.2cm}
\caption{\label{fig:Algorithm-block-diagram}The algorithmic framework of the
proposed CACRL, where the black line represents the flow for making
EEPS decisions online, and the orange line represents the flow for
network training.}
\vspace{-0.5cm}
\end{figure}
 To overcome the challenges in Remark 1, this section proposes a CACRL
algorithm, which consists of a CI module and a CRL module, as illustrated
in Fig. \ref{fig:Algorithm-block-diagram}.

The CI module integrates a context-aware meta-learning method and
a reward-reshaping method to infer $\bigl\{\boldsymbol{z}_{t}\bigr\}_{t=0,1,\ldots}$
and convert the original packet dropout rewards into immediate dense
ones, respectively. Specifically, following the context-aware meta-learning
method in \cite{VAE}, we assume that the latent variable $\boldsymbol{z}_{t}$
can be inferred from the most recent $N$ ($N\leq B$) transition
tuples $\boldsymbol{\tau}_{t}=\bigl\{\tilde{\tau}_{t'}\triangleq\bigl(\boldsymbol{s}_{t'},\boldsymbol{a}_{t'},\boldsymbol{s}_{t'+1}\bigr)\bigr\}_{t'=t-N+1:t}$
(which is the so-called \textit{context}) and train an encoder network
for this inference. Once $\boldsymbol{z}_{t}$ is obtained, it is
embed into $\boldsymbol{s}_{t}$ by
\begin{equation}
\dot{\boldsymbol{s}}_{t}\leftarrow\bigl\{\boldsymbol{s}_{t},\boldsymbol{z}_{t}\bigr\},\label{eq:embedding}
\end{equation}
where note that although the latent variable in (\ref{eq:embedding})
is merely an estimate of its true value obtained through the encoder
network, we still denote it as $\boldsymbol{z}_{t}$ for notational
simplicity. Differing from \cite{VAE}, the encoder network in this
paper is trained jointly with the CRL module in a model-free manner,
without relying on any actual or estimated knowledge of the environment
model, as elaborated in section \ref{subsec:The-CI-Module}. Since
the training of the CRL module depends on reward signals $\bigl\{ C_{k}\bigl(\boldsymbol{s}_{t},\boldsymbol{a}_{t}\bigr)\bigr\}_{t=0,1,\ldots},$$\forall k$,
sparse and delayed rewards could cause slow convergence or even failure
in these two modules. To overcome this challenge, the CI module also
trains some potential networks to construct a set of reward-reshaping
functions $\bigl\{ F_{\boldsymbol{\phi}_{k}}\bigr\}_{k=1:K}$ and
then transform the original packet dropout rewards into dense immediate
ones by 
\begin{equation}
\dot{C}_{k}^{'}\bigl(\dot{\boldsymbol{s}}_{t},\boldsymbol{a}_{t},\dot{\boldsymbol{s}}_{t+1}\bigr)\leftarrow C_{k}^{'}\bigl(\boldsymbol{s}_{t},\boldsymbol{a}_{t}\bigr)+F_{\boldsymbol{\phi}_{k,t}}\bigl(\dot{\boldsymbol{s}}_{t},\dot{\boldsymbol{s}}_{t+1}\bigr),\forall k,t,\label{eq:reshaping}
\end{equation}
where we abbreviate $\dot{C}_{k,t}^{'}$ for $\dot{C}_{k}^{'}\bigl(\dot{\boldsymbol{s}}_{t},\boldsymbol{a}_{t},\dot{\boldsymbol{s}}_{t+1}\bigr)$
in the following. In this way, effective optimizations can be performed
without waiting for the arrival of packet dropout rewards.

In the CRL module, $\pi_{\boldsymbol{\theta}}$ is chosen as the commonly
used Gaussian policy \cite{Gaussianpolicy}, whose mean and variance
are output by DNNs $\boldsymbol{\mu}_{\boldsymbol{\theta}_{\mu}}$
and $\mathbf{\Sigma}_{\boldsymbol{\theta}_{\sigma}}$, respectively,
where the parameter $\boldsymbol{\theta}=\bigl[\boldsymbol{\theta}_{\mu},\boldsymbol{\theta}_{\sigma}\bigr]\in\mathbf{\Theta}$.
In particular, to better handle non-convex stochastic constraints,
this paper employs a CSSCA-based policy optimization method. By employing
an Actor-Critic framework, the CRL module estimates the objective/constraint
values and gradients in (\ref{eq:CMDPs}) at the Critic step and further,
uses these estimates to construct a sequence of convex surrogate problems
at the Actor step. Then, the CSSCA-based policy optimization method
is executed once every $B$ ($N\leq B$) timeslots, to tackle the
original problem (\ref{eq:CMDPs}) by solving the sequence of surrogate
problems. After the $i$-th update, we obtain an observation set $\varepsilon_{i}=\bigl\{\tilde{\varepsilon}_{t}\bigr\}_{t=Bi-B+1:Bi}$,
where $\tilde{\varepsilon}_{t}\triangleq\bigl\{\dot{\boldsymbol{s}}_{t},\boldsymbol{a}_{t},\bigl\{\dot{C}_{k,t}^{'}\bigr\}_{k=0,\ldots,K},\dot{\boldsymbol{s}}_{t+1}\bigr\}$,
by interacting with the environment using $\pi_{\boldsymbol{\theta}_{i}}$,
and store it in a buffer. Note that the tuple $\tilde{\tau}_{t}$
in the CI module can be extracted from $\tilde{\varepsilon}_{t}$.

In practice, we treat the first $I$ iterations as pretraining for
the algorithm, and with the pretraining, the algorithm can work effectively
with the fine-tuned network at the deployment stage (after the $I$-th
iteration). Finally, the design details are provided below, and the
overall algorithm is summarized in Algorithm \ref{alg:CACRL-EEPS}.
\begin{algorithm}
\caption{\label{alg:CACRL-EEPS} The Context-Aware Constrained Reinforcement
Learning Algorithm}

\textbf{Input:} Initialize $\varrho_{1},\varrho_{2},\varrho_{3}$,
$\boldsymbol{\theta}_{0}$, $\boldsymbol{\psi}_{0}$, and $\bigl\{\boldsymbol{\omega}_{k,0},\boldsymbol{\phi}_{k,0}\bigr\}_{k=1:K}$.

\textbf{for} $i=0,1,\cdots$ \textbf{do}

\phantom{} \phantom{}\textbf{ }Sampling $\varepsilon_{i}$ based
on $\pi_{\boldsymbol{\theta}_{i}}$ and store it in the buffer.\vspace{0.1cm}

\phantom{} \phantom{}\textbf{ The CI Module:}

\phantom{} \phantom{} \textbf{for} $t=Bi-B+1,\cdots,Bi$ \textbf{do}

\phantom{} \phantom{} \phantom{} \phantom{} Calculate $E_{\boldsymbol{\psi}_{i}}\bigl(\boldsymbol{z}_{t}\bigr|\boldsymbol{\tau}_{t}\bigr)$
and sample $\boldsymbol{z}_{t}\sim E_{\boldsymbol{\psi}_{i}}\bigl(\boldsymbol{z}_{t}\bigr|\boldsymbol{\tau}_{t}\bigr)$.

\phantom{} \phantom{} \phantom{} \phantom{}\textbf{ }Replace $\boldsymbol{s}_{t}$
according to (\ref{eq:embedding})

\phantom{} \phantom{} \phantom{} \phantom{}\textbf{ }Calculate
$F_{\phi_{k,i}}\bigl(\boldsymbol{s}_{t},\boldsymbol{s}_{t+1}\bigr)=V_{\boldsymbol{\phi}_{k,i}}\bigl(\boldsymbol{s}_{t+1}\bigr)-V_{\boldsymbol{\phi}_{k,i}}\bigl(\boldsymbol{s}_{t}\bigr)$.

\phantom{} \phantom{} \phantom{} \phantom{} Reshape $\bigl\{ C_{k}^{'}\bigl(\boldsymbol{s}_{t},\boldsymbol{a}_{t}\bigr)\bigr\}_{k=1:K}$
by (\ref{eq:reshaping}).\vspace{0.1cm}

\phantom{} \phantom{}\textbf{ The CRL Module:}

\phantom{} \phantom{} \phantom{} \phantom{}\textbf{ }\textbf{\textit{The
Critic Step}}\textbf{:}

\phantom{} \phantom{} \phantom{} \phantom{}\textbf{ }Estimate
$\hat{f}_{k,i}$ according to (\ref{eq:f-hat}) and (\ref{eq:f-new}).

\phantom{} \phantom{} \phantom{} \phantom{}\textbf{ }Set $\boldsymbol{\omega}_{k,i}\leftarrow\boldsymbol{\omega}_{k,i-1}$,
$\boldsymbol{\psi}_{k,i}\leftarrow\boldsymbol{\psi}_{k,i-1}$, $\boldsymbol{\phi}_{k,i}\leftarrow\boldsymbol{\phi}_{k,i-1}$.

\phantom{} \phantom{} \phantom{} \phantom{} \textbf{for} $t_{\mathrm{cri}}=1,\cdots,T_{\mathrm{cri}}$
\textbf{do}

\phantom{} \phantom{}\textbf{ }\phantom{} \phantom{} \phantom{}
\phantom{}\textbf{ }Update $\boldsymbol{\omega}_{k,i}$, $\boldsymbol{\psi}_{k,i}$,
$\boldsymbol{\phi}_{k,i}$ by (\ref{eq:TD-updates}), (\ref{eq:update fi}),
(\ref{eq:V optimization}), respectively.

\phantom{} \phantom{} \phantom{} \phantom{}\textbf{ }Estimate
$\hat{\boldsymbol{g}}_{k,i}$ according to (\ref{eq:g-tilde}) and
(\ref{eq:g-hat}).

\phantom{} \phantom{} \phantom{} \phantom{}\textbf{ }\textbf{\textit{The
Actor Step}}\textbf{:}

\phantom{} \phantom{} \phantom{} \phantom{}\textbf{ }Update the
surrogate function $\left\{ \bar{f}_{k}\left(\boldsymbol{\theta}_{i}\right)\right\} _{k=0,\ldots,K}$via
(\ref{eq:surrogate functions}).

\phantom{} \phantom{} \phantom{} \phantom{} Solve (\ref{eq:objective update})
if (\ref{eq:objective update}) is feasible. \textit{(Objective update)}

\phantom{} \phantom{}\textbf{ }\phantom{} \phantom{} Solve (\ref{eq:feasible update})
if (\ref{eq:objective update}) is not feasible.\textit{ (Feasible
update)}

\phantom{} \phantom{} \phantom{} \phantom{}\textbf{ }Update policy
parameters $\boldsymbol{\theta}_{i+1}$ according to (\ref{eq:theta update}).
\end{algorithm}

\subsection{The CRL Module \label{subsec:The-CRL-Module}}

\subsubsection{The Actor Step Adopting CSSCA-based Policy Optimization}

At the $i$-th iteration, the CSSCA method replace the objective/constraint
functions $f_{k}\left(\boldsymbol{\theta}\right)$ by the following
convex quadratic surrogate functions based on $\boldsymbol{\theta}_{i}$:
\begin{equation}
\bar{f}_{k,i}\left(\boldsymbol{\theta}\right)=\hat{f}_{k,i}+\left(\hat{\boldsymbol{g}}_{k,i}\right)^{\textrm{\ensuremath{\intercal}}}\left(\boldsymbol{\theta}-\boldsymbol{\theta}_{i}\right)+\zeta_{k}\left\Vert \boldsymbol{\theta}-\boldsymbol{\theta}_{i}\right\Vert _{2}^{2},\forall k\label{eq:surrogate functions}
\end{equation}
where $\hat{f}_{k,i}\in\mathbb{R}$ and $\hat{\boldsymbol{g}}_{k,i}\in\mathbb{R}^{n_{\boldsymbol{\theta}}}$
are the esitimations of $f_{k}\left(\boldsymbol{\theta}\right)$ and
$\bigtriangledown f_{k}\left(\boldsymbol{\theta}\right)$, and $\zeta_{k}$
is a positive constant.

Then, based on these surrogate functions, the following objective
update problem is first solved:
\begin{align}
\bar{\boldsymbol{\theta}}_{i}=\underset{\boldsymbol{\theta}\in\mathbf{\Theta}}{\textrm{argmin}}\  & \bar{f}_{0,i}\left(\boldsymbol{\theta}\right)\ \ \ \ \textrm{s.t.}\ \bar{f}_{k,i}\left(\boldsymbol{\theta}\right)\leq0,\forall k.\label{eq:objective update}
\end{align}
If problem (\ref{eq:objective update}) turns out to be infeasible,
the following feasible update problem is solved: 
\begin{align}
\bar{\boldsymbol{\theta}}_{i}=\underset{\boldsymbol{\theta}\in\mathbf{\Theta},y}{\textrm{argmin}}\  & \alpha\ \ \ \ \textrm{s.t.}\ \bar{f}_{k,i}\left(\boldsymbol{\theta}\right)\leq\alpha,\forall k.\label{eq:feasible update}
\end{align}
Please note that the surrogate problems (\ref{eq:objective update})
and (\ref{eq:feasible update}) both belong to the convex quadratic
problem, which can be easily solved by standard convex optimization
algorithms, e.g. Lagrange-dual methods. Then, with $\bar{\boldsymbol{\theta}}_{i}$
given in one of the above two cases, $\boldsymbol{\theta}_{i}$ is
updated according to
\begin{equation}
\boldsymbol{\theta}_{i}=(1-\mu_{i})\boldsymbol{\theta}_{i-1}+\mu_{i}\bar{\boldsymbol{\theta}}_{i}.\label{eq:theta update}
\end{equation}
where the step size $\bigl\{\mu_{i}\bigr\}$ is a decreasing sequence
satisfying Assumption 2 in section \ref{subsec:Key Assumptions}.

\subsubsection{The Critic Step for Estimating Function Values and Gradients}

The Critic step aims to calculate the estimated function values $\bigl\{\hat{f}_{k,i+1}\bigr\}_{k=0,\ldots,K}$
and the estimated gradients $\bigl\{\hat{\boldsymbol{g}}_{k,i+1}\bigr\}_{k=0,\ldots,K}$
based on their historical values and the new observation set $\varepsilon_{i}$,
and then uses them to construct $\bigl\{\bar{f}_{k,i+1}\left(\boldsymbol{\theta}\right)\bigr\}_{k=0,\ldots,K}$
in the next iteration. First, we calculate a new estimation of function
value using the sample average approximation (SAA) method as follows:
\begin{equation}
\tilde{f}_{k,i}=\hat{\mathbb{E}}_{\varepsilon_{i}}\bigl[\dot{C}_{k,t}^{'}\bigr],\forall k,\label{eq:f-new}
\end{equation}
where $\hat{\mathbb{E}}_{\varepsilon_{i}}$ denotes the sample average
operation over the observation set $\varepsilon_{i}$. To reduce oscillations
and accelerate convergence, we update $\hat{f}_{k,i+1}$ by
\begin{equation}
\hat{f}_{k,i+1}=\left(1-\eta_{i+1}\right)\hat{f}_{k,i}+\eta_{i}\tilde{f}_{k,i},\forall k,\label{eq:f-hat}
\end{equation}
where the step size $\bigl\{\eta_{i}\bigr\}$ is a decreasing sequence
satisfying Assumption 2 in section \ref{subsec:Key Assumptions}.

Then, for any $\pi_{\boldsymbol{\theta}_{i}}$, there is a set of
state-action value functions (Q-functions) $\bigl\{ Q_{k}^{\pi_{\boldsymbol{\theta}_{i}}}\bigr\}_{k=0,\ldots,K}$
formulated as:
\begin{align*}
Q_{k}^{\pi_{\boldsymbol{\theta}_{i}}}\left(\dot{\boldsymbol{s}},\boldsymbol{a}\right)= & \mathbb{E}\Bigl[\sum_{l=0}^{\infty}\Bigl(\dot{C}_{k,l}^{\text{'}}-f_{k,i}\left(\boldsymbol{\theta}\right)\Bigr)\bigl|\dot{\boldsymbol{s}}_{0},\boldsymbol{a}_{0}=\dot{\boldsymbol{s}},\boldsymbol{a}\Bigr].
\end{align*}
Since it is unrealistic to obtain $f_{k,i}\left(\boldsymbol{\theta}\right)$
online, we replace it by $\hat{f}_{k,i}$ and redefine a surrogate
Q-function of $Q_{k}^{\pi_{\boldsymbol{\theta}_{i}}},\forall k$:
\begin{align*}
\hat{Q}_{k}^{\pi_{\boldsymbol{\theta}_{i}}}\left(\dot{\boldsymbol{s}},\boldsymbol{a}\right)= & \mathbb{E}\Bigl[\sum_{l=0}^{\infty}\Bigl(\dot{C}_{k,l}^{\text{'}}-\hat{f}_{k,i}\Bigr)\bigl|\dot{\boldsymbol{s}}_{0},\boldsymbol{a}_{0}=\dot{\boldsymbol{s}},\boldsymbol{a}\Bigr].
\end{align*}
Considering the state space $\mathcal{S}$ and the action space $\mathcal{A}$
are continuous, a set of DNNs (Q-networks) $\bigl\{ Q_{\boldsymbol{\omega}_{k,i}}\bigr\}_{k=0,\ldots,K}$
is employed to parameterize $\bigl\{\hat{Q}_{k}^{\pi_{\boldsymbol{\theta}_{i}}}\bigr\}_{k=0,\ldots,K}$.
The gap between the estimated Q-value $Q_{\boldsymbol{\omega}_{k,i}}\left(\dot{\boldsymbol{s}},\boldsymbol{a}\right)$
and the Q-value $\hat{Q}_{k}^{\pi_{\boldsymbol{\theta}_{i}}}\left(\dot{\boldsymbol{s}},\boldsymbol{a}\right)$
can be measured using the squared Bellman error:
\begin{align}
\mathcal{B}_{k,i}\bigl(\dot{\boldsymbol{s}}_{t},\boldsymbol{a}_{t},\dot{\boldsymbol{s}}_{t+1}\bigr)= & \bigl|Q_{\boldsymbol{\omega}_{k,i}}\left(\dot{\boldsymbol{s}}_{t},\boldsymbol{a}_{t},\dot{\boldsymbol{s}}_{t+1}\right)\label{eq:Bellman error}\\
 & -\mathcal{T}Q_{\boldsymbol{\omega}_{k,i}}\left(\dot{\boldsymbol{s}}_{t},\boldsymbol{a}_{t},\dot{\boldsymbol{s}}_{t+1}\right)\bigr|^{2},\forall k,\nonumber 
\end{align}
where $\mathcal{T}$ is the Bellman operator defined as
\[
\mathcal{T}Q_{\boldsymbol{\omega}_{k,i}}\left(\dot{\boldsymbol{s}}_{l},\boldsymbol{a}_{l}\right)=\mathbb{E}\Bigl[Q_{\boldsymbol{\omega}_{k,i}}\left(\dot{\boldsymbol{s}}_{l+1},\boldsymbol{a}_{l+1}'\right)+\dot{C}_{k,l}^{\text{'}}\Bigr]-\hat{f}_{k,i}.
\]
The Q-networks are optimized to minimize the mean-squared Bellman
error using the classical TD-learning method \cite{SCAOPO}. Specifically,
the new observation set $\varepsilon_{i}$ is divided into $T_{\mathrm{cri}}$
($T_{\mathrm{cri}}<B$) batches, where the observation index set of
these batches are $\bigl[\varepsilon_{i}^{1},\ldots,\varepsilon_{i}^{t_{\mathrm{cri}}},\ldots,\varepsilon_{i}^{T_{\mathrm{cri}}}\bigr]$,
and the following TD-updates are performed for $T_{\mathrm{cri}}$
times:
\begin{equation}
\boldsymbol{\omega}_{k,i}\leftarrow\boldsymbol{\omega}_{k,i}-\upsilon_{i}G_{k,i,t_{\mathrm{cri}}}^{\mathcal{B}},\forall k,t_{\mathrm{cri}},\label{eq:TD-updates}
\end{equation}
where $\boldsymbol{G}_{k,i,t_{\mathrm{cri}}}^{\mathcal{B}}$ is a
stochastic gradient term given by
\begin{align}
\boldsymbol{G}_{k,i,t_{\mathrm{cri}}}^{\mathcal{B}}= & \sum_{t\in\varepsilon_{i}^{t_{\mathrm{cri}}}}\Bigl(Q_{\boldsymbol{\omega}_{k,i}}\left(\dot{\boldsymbol{s}}_{t},\boldsymbol{a}_{t}\right)-\dot{C}_{k,t}^{\text{'}}+\hat{f}_{k,i}\nonumber \\
-Q_{\boldsymbol{\omega}_{k,i}} & \bigl(\dot{\boldsymbol{s}}_{t+1},\boldsymbol{a}_{t+1}'\bigr)\Bigr)\nabla_{\boldsymbol{\omega}}Q_{\boldsymbol{\omega}_{k,i}}\left(\dot{\boldsymbol{s}}_{t},\boldsymbol{a}_{t}\right),\forall k,\label{eq:gradient w}
\end{align}
where the step size $\bigl\{\upsilon_{i}\bigr\}$ is a decreasing
sequence satisfying Assumption 2 in section \ref{subsec:Key Assumptions}.

Further, according to the policy gradient theorem \cite{DQlearning}:
\[
\nabla f_{k,i}\left(\boldsymbol{\theta}\right)=\mathbb{E}\left[Q_{k}^{\pi_{\boldsymbol{\theta}_{i}}}\left(\dot{\boldsymbol{s}},\boldsymbol{a}\right)\nabla_{\boldsymbol{\theta}}\textrm{log}\pi_{\boldsymbol{\theta}_{i}}\left(\boldsymbol{a}\mid\dot{\boldsymbol{s}}\right)\right],\forall k,
\]
and following the SAA method, we calculate the new estimated gradient
$\tilde{\boldsymbol{g}}_{k,i}$ by
\begin{align}
\tilde{\boldsymbol{g}}_{k,i}= & \hat{\mathbb{E}}_{\varepsilon_{i}}\left[Q_{\boldsymbol{\omega}_{k,i}}\left(\dot{\boldsymbol{s}},\boldsymbol{a}\right)\nabla_{\boldsymbol{\theta}}\textrm{log}\pi_{\boldsymbol{\theta}_{i}}\left(\boldsymbol{a}\mid\dot{\boldsymbol{s}}\right)\right],\forall k.\label{eq:g-tilde}
\end{align}
Finally, similar to (\ref{eq:f-hat}), $\hat{\boldsymbol{g}}_{k,i}$
is updated according to
\begin{equation}
\hat{\boldsymbol{g}}_{k,i+1}=\left(1-\eta_{i+1}\right)\hat{\boldsymbol{g}}_{k,i}+\eta_{i}\tilde{\boldsymbol{g}}_{k,i},\forall k.\label{eq:g-hat}
\end{equation}

\subsection{The CI Module \label{subsec:The-CI-Module}}

\subsubsection{Context-aware Meta-learning}

Due to the potential variation of latent variables at each timeslot,
the CI module estimates $\boldsymbol{z}_{t}$ through a $\boldsymbol{\psi}$-parameterized
encoder network $E_{\boldsymbol{\psi}}$ at each timeslot $t$. Specifically,
we use $E_{\boldsymbol{\psi}}\bigl(\boldsymbol{z}_{t}\bigr|\boldsymbol{\tau}_{t}\bigr)$
to approximate the posterior of $\boldsymbol{z}_{t}$ and sample $\boldsymbol{z}_{t}$
from $E_{\boldsymbol{\psi}}\bigl(\boldsymbol{z}_{t}\bigr|\boldsymbol{\tau}_{t}\bigr)$,
where recall that $\boldsymbol{\tau}_{t}=\bigl\{\tilde{\tau}_{t'}=\bigl(\boldsymbol{s}_{t'},\boldsymbol{a}_{t'},\boldsymbol{s}_{t'+1}\bigr)\bigr\}_{t'=t-N+1:t}$.
Note that this probabilistic modeling of $\boldsymbol{z}_{t}$ can
enable CRL algorithms to explore more efficiently, due to the introduction
of extra randomness \cite{PERAL}. Considering the encoder for a CMDP
should be independent of the permutation of context tuples, its structure
is designed as a product of some independent factors:\vspace{-0.2cm}
\[
E_{\boldsymbol{\psi}}\bigl(\boldsymbol{z}_{t}\bigr|\boldsymbol{\tau}_{t}\bigr)\propto\prod_{t'=t-N+1}^{N}e_{\boldsymbol{\psi}}\bigl(\boldsymbol{z}_{t'}\bigr|\tilde{\boldsymbol{\tau}}_{t'}\bigr)=\mathcal{N}\Bigl(E_{\boldsymbol{\psi}}^{u}\bigl(\boldsymbol{\tau}_{t}\bigr),E_{\boldsymbol{\psi}}^{\sigma}\bigl(\boldsymbol{\tau}_{t}\bigr)\Bigr),
\]
where $e_{\boldsymbol{\psi}}\bigl(\boldsymbol{z}_{t}\bigr|\tilde{\tau}_{t'}\bigr)$
represents the Gaussian distribution $\mathcal{N}\bigl(e_{\boldsymbol{\psi}_{\boldsymbol{\mu}}}^{u}\bigl(\tilde{\tau}_{t'}\bigr),e_{\boldsymbol{\psi_{\sigma}}}^{\sigma}\bigl(\tilde{\tau}_{t'}\bigr)\bigr)$
with $\boldsymbol{\psi}\triangleq\bigl[\boldsymbol{\psi}_{\mu},\boldsymbol{\psi}_{\sigma}\bigr]$.
Moreover, $E_{\boldsymbol{\psi}}^{u}\bigl(\boldsymbol{\tau}_{t}\bigr)$
and $E_{\boldsymbol{\psi}}^{\sigma}\bigl(\boldsymbol{\tau}_{t}\bigr)$
can be respectively given by\vspace{-0.2cm}
\[
E_{\boldsymbol{\psi}}^{u}\bigl(\boldsymbol{\tau}_{t}\bigr)=E_{\boldsymbol{\psi}}^{\sigma}\bigl(\boldsymbol{\tau}_{t}\bigr)\cdot\prod_{t'=t-N+1}^{t}e_{\boldsymbol{\psi}_{\boldsymbol{\mu}}}^{u}\bigl(\tilde{\tau}_{t'}\bigr)/e_{\boldsymbol{\psi_{\sigma}}}^{\sigma}\bigl(\tilde{\tau}_{t'}\bigr),
\]
\vspace{-0.2cm}
\[
E_{\boldsymbol{\psi}}^{\sigma}\bigl(\boldsymbol{\tau}_{t}\bigr)=\Bigl(\prod_{t'=t-N+1}^{t}\frac{1}{e_{\boldsymbol{\psi_{\sigma}}}^{\sigma}\bigl(\tilde{\boldsymbol{\tau}}_{t'}\bigr)}\Bigr)^{-1}.
\]

Now, we elaborate on how to optimize the encoder. Previous work \cite{VAE}
directly uses a $\boldsymbol{z}$-parameterized DNN to fit the transition
function and optimizes $\boldsymbol{z}$ to reconstruct to reconstruct
trajectory $\bigl\{\boldsymbol{s}_{0},\boldsymbol{a}_{0},\boldsymbol{s}_{1},\boldsymbol{a}_{1},\ldots\bigr\}$,
which belongs to a model-based optimization method. However, this
method incurs substantial additional training costs, as estimating
the transition function is unnecessary in standard model-free Actor-Critic
algorithms, and model-based approaches often yield significant errors
in the learned transition function when the environment model is complex
\cite{modelbased}. Considering that $\hat{Q}_{k}^{\pi_{\boldsymbol{\theta}}}$
is defined as the future expected average reward over the current
policy and transition probability function $P_{\mathrm{T}}$, it is
clear that the optimization process of Q-networks involves learning
$P_{\mathrm{T}}$ in a model-free manner. This motivates this paper
to optimizes $E_{\boldsymbol{\psi}}\bigl(\boldsymbol{z}\bigr|\boldsymbol{\tau}\bigr)$
and Q-networks $\bigl\{ Q_{\boldsymbol{\omega}_{k}}\bigr\}_{k=0,\ldots,K}$
simultaneously, with the goal of enabling $\boldsymbol{z}$ to aid
$\bigl\{ Q_{\boldsymbol{\omega}_{k}}\bigr\}_{k=0,\ldots,K}$ in reconstructing
the Q-functions $\bigl\{\hat{Q}_{k}^{\pi_{\boldsymbol{\theta}}}\bigr\}_{k=0,\ldots,K}$,
where more accurate $\boldsymbol{z}$ estimates lead to smaller reconstruction
errors.

To facilitate joint training with the CRL module, the encoder network
is updated once every $B$ timeslots. Specifically, at the $i$-th
iteration, we first define an optimal fitting variable $O_{i}$ with
$O_{i}=1$ representing the Q-networks $\bigl\{ Q_{\boldsymbol{\omega}_{k,i}}\bigr\}_{k=0,\ldots,K}$
achieves optimal reconstruction for the Q-functions $\bigl\{\hat{Q}_{k}^{\pi_{\boldsymbol{\theta}}}\bigr\}_{k=0,\ldots,K}$.
Accordingly, $P\bigl(O_{i}=1\bigr|\boldsymbol{\boldsymbol{z}}_{i}\bigr)$
represents the probability that $O_{i}=1$ holds given $\boldsymbol{\boldsymbol{z}}_{i}=\left\{ \boldsymbol{z}_{t},Bi-N+1\leq t\leq Bi\right\} $.
Considering that the squared Bellman errors $\mathcal{B}_{k,i}$ in
(\ref{eq:Bellman error}) reflect the reconstruction error of Q-functions,
we related it to $P\bigl(O_{i}=1\bigr|\boldsymbol{\boldsymbol{z}}_{i}\bigr)$
by
\begin{align*}
P\bigl(O_{i}=1\bigr|\boldsymbol{\boldsymbol{z}}_{i}\bigr)\propto\mathrm{exp} & \Bigl(-\sum_{t=Bi-B+1}^{Bi}\sum_{k=1}^{K}\mathcal{B}_{k,i}\bigl(\dot{\boldsymbol{s}}_{t},\boldsymbol{a}_{t},\dot{\boldsymbol{s}}_{t+1}\bigr)\Bigr),
\end{align*}
Moreover, $P\bigl(\boldsymbol{\boldsymbol{z}}_{i}\bigr|O_{i}=1\bigr)$
represents the probability that $\boldsymbol{\boldsymbol{z}}_{i}$
is the most effective for guiding the reconstruction of Q-functions
during the $i$-th interation. Following the mean-field varational
Bayesian inference method \cite{meanfield}, we use a varational distribution
of product form $q\left(\boldsymbol{z}_{i};\boldsymbol{\psi}\right)=\prod_{t=Bi-B+1}^{Bi}E_{\boldsymbol{\psi}}\bigl(\boldsymbol{z}_{t}\bigr|\boldsymbol{\tau}_{t}\bigr)$
to approximate $P\bigl(\boldsymbol{\boldsymbol{z}}_{i}\bigr|O_{i}=1\bigr)$.
This optimization objective is to minimize the Kullback-Leibler Divergence
between $q\left(\boldsymbol{z}_{i};\boldsymbol{\psi}\right)$ and
$P\bigl(\boldsymbol{\boldsymbol{z}}_{i}\bigr|O_{i}=1\bigr)$ as:\vspace{-0.2cm}
\begin{align}
 & \sum_{i=0}^{\infty}D_{\mathrm{KL}}\Bigl[q\left(\boldsymbol{z}_{i};\boldsymbol{\psi}\right)||P\bigl(\boldsymbol{z}_{i}\bigr|O_{i}=1\bigr)\Bigr]\label{eq:ELBO}\\
= & \sum_{i=0}^{\infty}\mathbb{E}_{\boldsymbol{z}_{t}\sim E_{\boldsymbol{\psi}}}\biggl[-\mathrm{log}\frac{P\bigl(O_{i}=1\bigr|\boldsymbol{z}_{i}\bigr)P\bigl(\boldsymbol{z}_{i}\bigr)}{q\left(\boldsymbol{z}_{i};\boldsymbol{\psi}\right)P\bigl(O_{i}=1\bigr)}\biggr]\nonumber \\
\propto & \mathbb{E}_{\boldsymbol{z}_{t}\sim E_{\boldsymbol{\psi}}}\Bigl[-\mathrm{log}P\bigl(O_{i}=1\bigr|\boldsymbol{z}_{i}\bigr)\Bigr]+D_{\mathrm{KL}}\Bigl[q\left(\boldsymbol{z}_{i};\boldsymbol{\psi}\right)||P\bigl(\boldsymbol{z}_{i}\bigr)\Bigr]\nonumber \\
= & \underset{\mathcal{L}_{\mathrm{meta}}^{1}\bigl(\boldsymbol{\psi}\bigr)}{\underbrace{\sum_{i=0}^{\infty}\mathbb{E}_{\boldsymbol{z}_{t}\sim E_{\boldsymbol{\psi}}}\Bigl[\sum_{t=Bi-B+1}^{Bi}\sum_{k=1}^{K}\mathcal{B}_{k,i}\bigl(\dot{\boldsymbol{s}}_{t},\boldsymbol{a}_{t},\dot{\boldsymbol{s}}_{t+1}\bigr)\Bigr]}}\nonumber \\
 & \underset{\mathcal{L}_{\mathrm{meta}}^{2}\bigl(\boldsymbol{\psi}\bigr)}{+\underbrace{\sum_{i=0}^{\infty}\sum_{t=Bi-B+1}^{Bi}D_{\mathrm{KL}}\Bigl[E_{\boldsymbol{\psi}}\bigl(\boldsymbol{z}_{t}\bigr|\boldsymbol{\tau}_{t}\bigr)||P\bigl(\boldsymbol{z}_{t}\bigr)\Bigr]}},\nonumber 
\end{align}

We solve this by performing the following SGD for $T_{\mathrm{cri}}$
times at the $i$-th iteration:\vspace{-0.2cm}
\begin{equation}
\boldsymbol{\psi}_{i}\leftarrow\boldsymbol{\psi}_{i}-\upsilon_{i}\Bigl(\sum_{k=1}^{K}\boldsymbol{G}_{k,i,t_{\mathrm{cri}}}^{\mathcal{L}^{1}}+\boldsymbol{G}_{i,t_{\mathrm{cri}}}^{\mathcal{L}^{2}}\Bigr),\forall t_{\mathrm{cri}},\label{eq:update fi}
\end{equation}
where $\boldsymbol{G}_{k,i,t_{\mathrm{cri}}}^{\mathcal{L}}$ and $\boldsymbol{G}_{i,t_{\mathrm{cri}}}^{\mathcal{L}^{2}}$
are stochastic gradients of $\mathcal{L}_{\mathrm{meta}}^{1}\bigl(\boldsymbol{\psi}\bigr)$
and $\mathcal{L}_{\mathrm{meta}}^{2}\bigl(\boldsymbol{\psi}\bigr)$.
The term $G_{k,i,t_{\mathrm{cri}}}^{\mathcal{L}^{1}}$ is given by
\begin{align*}
\boldsymbol{G}_{k,i,t_{\mathrm{cri}}}^{\mathcal{L}^{1}}= & \sum_{t\in\varepsilon_{i}^{t_{\mathrm{cri}}}}\Bigl(Q_{\boldsymbol{\omega}_{k,i}}\left(\dot{\boldsymbol{s}}_{t},\boldsymbol{a}_{t}\right)-C_{k,t}^{\text{'}}+\hat{f}_{k,i}\\
-Q_{\boldsymbol{\omega}_{k,i}}\bigl( & \dot{\boldsymbol{s}}_{t+1},\boldsymbol{a}_{t+1}'\bigr)\Bigr)\nabla_{\boldsymbol{z}_{t}}Q_{\boldsymbol{\omega}_{k,i}}\left(\dot{\boldsymbol{s}}_{t},\boldsymbol{a}_{t}\right)\cdot\nabla_{\boldsymbol{\psi}}\boldsymbol{z}_{t}
\end{align*}
where denoting $\xi_{t_{\mathrm{cri}}}$ as a constant generated from
Gaussian distribution $\mathcal{N}\bigl(0,1\bigr)$, we have
\begin{align*}
\nabla_{\boldsymbol{\psi}}\boldsymbol{z}_{t}= & \nabla_{\boldsymbol{\psi}}E_{\boldsymbol{\psi}}^{u}\bigl(\boldsymbol{\tau}_{t}\bigr)+\xi_{t}\nabla_{\boldsymbol{\psi}}E_{\boldsymbol{\psi}}^{\sigma}\bigl(\boldsymbol{\tau}_{t}\bigr).
\end{align*}
Note that $\boldsymbol{G}_{k,i,t_{\mathrm{cri}}}^{\mathcal{L}^{1}}$
can be calculated simultaneously with $\boldsymbol{G}_{k,i,t_{\mathrm{cri}}}^{\mathcal{B}}$
in (\ref{eq:gradient w}) simultaneously through the same backpropagations.
Moreover, the stochastic gradient $\boldsymbol{G}_{i,t_{\mathrm{cri}}}^{\mathcal{L}^{2}}$
is given by
\begin{align*}
\boldsymbol{G}_{i,t_{\mathrm{cri}}}^{\mathcal{L}^{2}}=\sum_{t\in\varepsilon_{i}^{t_{\mathrm{cri}}}} & E_{\boldsymbol{\psi}}^{u}\bigl(\boldsymbol{\tau}_{t}\bigr)\nabla_{\boldsymbol{\psi}}E_{\boldsymbol{\psi}}^{u}\bigl(\boldsymbol{\tau}_{t}\bigr)-\frac{\nabla_{\boldsymbol{\psi}}E_{\boldsymbol{\psi}}^{\sigma}\bigl(\boldsymbol{\tau}_{t}\bigr)}{2E_{\boldsymbol{\psi}}^{\sigma}\bigl(\boldsymbol{\tau}_{t}\bigr)}.
\end{align*}

\subsubsection{Potential-based Reward-Reshaping}

The reward-reshaping functions in (\ref{eq:reshaping}) are chosen
to be $F_{\boldsymbol{\phi}_{k,i}}\bigl(\dot{\boldsymbol{s}}_{t},\dot{\boldsymbol{s}}_{t+1}\bigr)=V_{\boldsymbol{\phi}_{k,i}}\bigl(\dot{\boldsymbol{s}}_{t+1}\bigr)-V_{\boldsymbol{\phi}_{k,i}}\bigl(\dot{\boldsymbol{s}}_{t}\bigr),k=1,\ldots,K$,
where $V_{\boldsymbol{\phi}_{k,i}}$ is a $\boldsymbol{\phi}_{k,i}$-parameterized
potential networks used to approximate the following optimal value
function \cite{reward0shaping1}:
\begin{equation}
V_{k}^{\pi_{\boldsymbol{\theta}^{*}}}\bigl(\dot{\boldsymbol{s}}\bigr)=\mathbb{E}_{\boldsymbol{a}\sim\pi_{\boldsymbol{\theta}^{*}}}\bigl[Q_{k}^{\pi_{\boldsymbol{\theta}^{*}}}\bigl(\dot{\boldsymbol{s}},\boldsymbol{a}\bigr)\bigr],\forall k,\label{eq:definitaion V}
\end{equation}
where $\boldsymbol{\theta}^{*}$ is a KKT point of problem (\ref{eq:CMDPs}).
Note that since the input space of $V_{k}^{\pi_{\boldsymbol{\theta}^{*}}}$
is much smaller than that of $Q_{k}^{\pi_{\boldsymbol{\theta}^{*}}}$,
it is also easier to be estimated and can serve as the prior representing
the risk of packet dropout.
\begin{figure}[h]
\centering{}\vspace{-0.3cm}
\includegraphics[width=5cm,height=1.7cm]{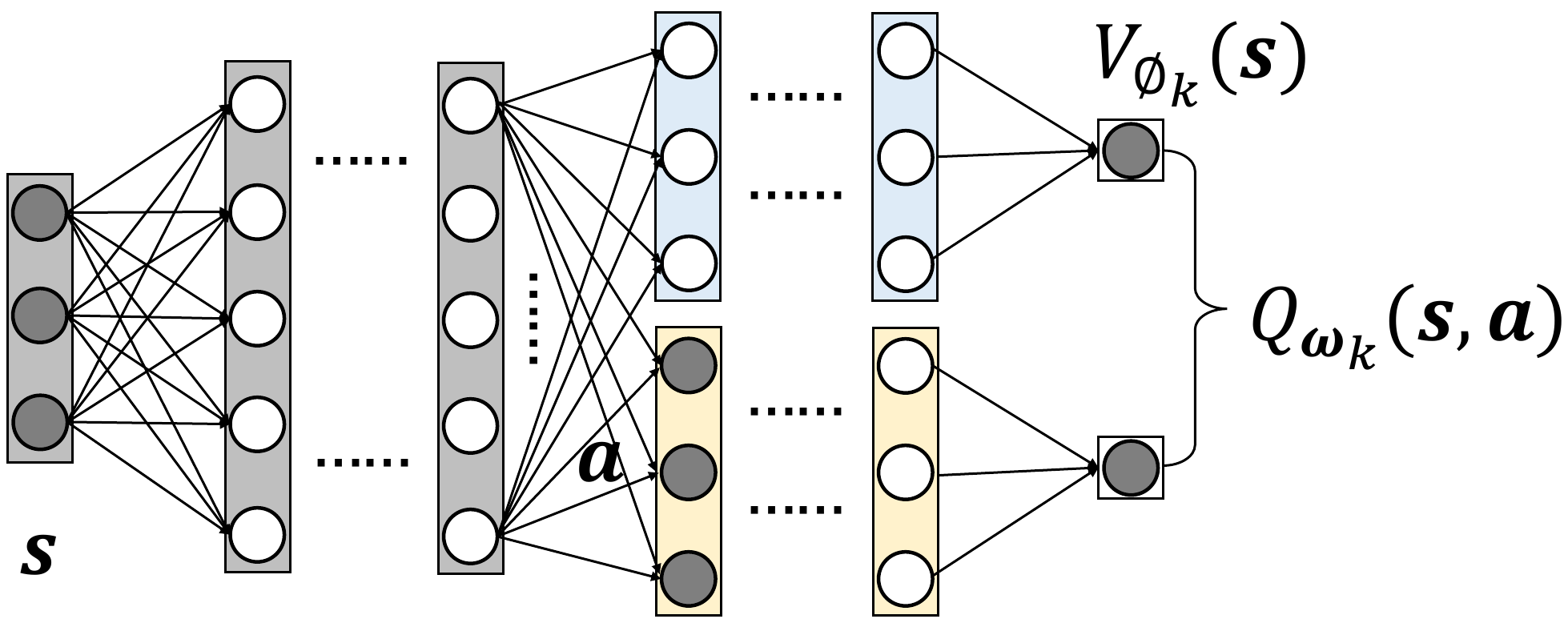}\vspace{-0.3cm}
\caption{\label{fig:dual-head network}The framework of dual-head networks}
\vspace{-0.4cm}
\end{figure}

Then, the potential networks $\bigl\{ V_{\boldsymbol{\phi}_{k}}\bigr\}_{k=1,\ldots,K}$
are also optimized with Q-networks $\bigl\{ Q_{\boldsymbol{\omega}_{k}}\bigr\}_{k=1,\ldots,K}$
simultaneously. To reduce computational costs, we integrate them into
$K$ dual-head networks shown in Fig. \ref{fig:dual-head network},
whose shallow layers are shared for extracting features from $\dot{\boldsymbol{s}}_{t}$.
According to (\ref{eq:definitaion V}), we optimize $\bigl\{ V_{\boldsymbol{\phi}_{k,i}}\bigr\}_{k=1,\ldots,K}$
by performing the following SGD for $T_{\mathrm{cri}}$ times at the
$i$-th iteration:
\begin{align}
\boldsymbol{\phi}_{k,i}\leftarrow & \boldsymbol{\phi}_{k,i}-\upsilon_{i}\sum_{t\in\varepsilon_{i}^{t_{\mathrm{cri}}}}\bigl[\nabla_{\boldsymbol{\phi}}\bigl|V_{\boldsymbol{\phi}_{k,i}}\bigl(\dot{\boldsymbol{s}}_{t}\bigr)-\hat{V}_{k,t}\bigr|^{2}\bigr],\forall k,\label{eq:V optimization}
\end{align}
where $\hat{V}_{k,t}=\frac{1}{N_{a}}\sum_{n_{a}=1}^{N_{a}}Q_{\boldsymbol{\omega}_{k,i}}\bigl(\dot{\boldsymbol{s}}_{t},\boldsymbol{a}_{t}^{n_{a}}\bigr)$
is the SAA-based estimation of $\mathbb{E}_{\boldsymbol{a}\sim\pi_{\boldsymbol{\theta}}}\bigl[Q^{\pi_{\boldsymbol{\theta}_{i}}}\bigl(\dot{\boldsymbol{s}},\boldsymbol{a}\bigr)\bigr]$
with $\bigl\{\boldsymbol{a}^{n_{a}}\bigr\}_{n_{a}=1:N_{a}}$ representing
$N_{a}$ actions sampled from $\pi_{\boldsymbol{\theta}_{i}}\bigl(\cdot\mid\dot{\boldsymbol{s}}\bigr)$,
and $\hat{V}_{k,i}$ is treated as a variable independent of $\boldsymbol{\phi}_{k,i}$
in the derivation operation.

\section{Theoretical Analyses of Algorithm Convergence\label{sec:Theoretical-Analyses}}

To provide deeper insight into the CACRL algorithm, this section presents
some theoretical analyses in an ideal case that CI module can accurately
estimate $\bigl\{\boldsymbol{z}_{t}\bigr\}_{t=0,1,\ldots}$ and $V_{k}^{\pi_{\boldsymbol{\theta}^{*}}}\bigl(\dot{\boldsymbol{s}}\bigr)$
through the encoder and potential networks. Specifically, we first
prove that reward-reshaping does not change the optimal policy of
problem (\ref{eq:CMDPs}). Then, based on some standard assumptions,
we analyze the convergence of surrogate functions in the CRL module
and prove that the CACRL can converge to a KKT point (up to a tolerable
error) of problem (\ref{eq:CMDPs}).

\subsection{Optimal Policy Invariance of the Reward-Reshaping \label{subsec:Optimal-Policy-Invariance}}

By performing (\ref{eq:reshaping}), the reward-reshaping method transforms
the original problem (\ref{eq:CMDPs}) into the following problem:\vspace{-0.2cm}
\begin{align}
\underset{\theta\in\Theta}{\mathrm{min}}\dot{f}_{0}\left(\boldsymbol{\theta}\right) & \thinspace\thinspace\thinspace\thinspace\thinspace\mathrm{s.t.}\dot{f}_{k}\left(\boldsymbol{\theta}\right),k=1,\ldots,K,\label{eq:CMDPs-shaped}
\end{align}
where $\bigl\{\dot{f}_{k}\left(\boldsymbol{\theta}\right)\bigr\}_{k=0,\ldots,K}$
are reshaped functions given by replacing $\bigl\{ C_{k}^{'}\bigr\}_{k=1,\ldots,K}$
in the definition of $\bigl\{ f_{k}\left(\boldsymbol{\theta}\right)\bigr\}_{k=0,\ldots,K}$
with $\bigl\{\dot{C}_{k}^{'}\bigr\}_{k=1,\ldots,K}$. By deriving
the expression for the reshaped function $\dot{f}_{k}\left(\boldsymbol{\theta}\right)$
, we further obtain\vspace{-0.2cm}
\begin{align}
\dot{f}_{k} & \left(\boldsymbol{\theta}\right)=\lim_{T\rightarrow\infty}\frac{1}{T}\mathbb{E}\Bigl[\stackrel[t=0]{T-1}{\sum}C_{k}^{'}\left(\boldsymbol{s}_{t},\boldsymbol{a}_{t}\right)+F_{\boldsymbol{\phi}_{k}}\bigl(\dot{\boldsymbol{s}}_{t},\dot{\boldsymbol{s}}_{t+1}\bigr)\Bigr],\nonumber \\
= & \underset{T\rightarrow\infty}{\mathrm{lim}}\frac{1}{T}\mathbb{E}\bigl[\stackrel[t=0]{T-1}{\sum}C_{k}^{'}\left(\boldsymbol{s}_{t},\boldsymbol{a}_{t}\right)+V_{\boldsymbol{\phi}_{k}}\bigl(\dot{\boldsymbol{s}}_{t+1}\bigr)-V_{\boldsymbol{\phi}_{k}}\bigl(\dot{\boldsymbol{s}}_{t}\bigr)\bigr],\nonumber \\
= & \underset{T\rightarrow\infty}{\mathrm{lim}}\frac{1}{T}\bigl[V_{\boldsymbol{\phi}_{k}}\bigl(\dot{\boldsymbol{s}}_{0}\bigr)-V_{\boldsymbol{\phi}_{k}}\bigl(\dot{\boldsymbol{s}}_{T-1}\bigr)\bigr]\nonumber \\
 & +\underset{T\rightarrow\infty}{\mathrm{lim}}\frac{1}{T}\mathbb{E}\bigl[\stackrel[t=0]{T-1}{\sum}C_{k}^{'}\left(\boldsymbol{s}_{t},\boldsymbol{a}_{t}\right)\bigr]\nonumber \\
= & f_{k}\left(\boldsymbol{\theta}\right),k.
\end{align}
Therefore, we can conclude that the reward-reshaping method does not
change the optimal policy of the problem (\ref{eq:CMDPs}).

\subsection{Key Assumptions on Problem and Step Sizes \label{subsec:Key Assumptions}}

We first give some assumptions on the problem structure:

\noindent \newtheorem{assumption}{Assumption}
\begin{assumption}\textit{ (Assumptions on the Problem Structure:)}\\
1) The state space $\mathcal{S}\subseteq\mathbb{R}^{n_{\boldsymbol{s}}}$,
the latent variable space $\mathcal{Z}\subseteq\mathbb{R}^{n_{\boldsymbol{z}}}$,
and the action space $\mathcal{A}\subseteq\mathbb{R}^{n_{\boldsymbol{a}}}$
are compact, where $n_{\boldsymbol{s}}$, $n_{\boldsymbol{z}}$, and
$n_{\boldsymbol{a}}$ are some positive integers. The functions $\bigl\{\dot{C}_{k}^{'}\bigr\}_{k=0,\ldots,K}$,
are bounded.\\
2) The policy $\pi_{\boldsymbol{\theta}}$ follows Lipschitz continuity
over $\boldsymbol{\theta}\in\mathbf{\Theta}$, where the parameter
spaces $\mathbf{\Theta}\subseteq\mathbb{R}^{n_{\boldsymbol{\theta}}}$
is compact and convex.\\
3) The gradients of $\dot{f}_{k}\left(\boldsymbol{\theta}\right),\forall k$,
are uniformly bounded follow Lipschitz continuity over the parameter
$\boldsymbol{\theta}\in\mathbf{\Theta}$.\\
4) There are constants $\lambda>0$ and $\rho\in\left(0,1\right)$
satisfying
\[
\mathrm{sup}_{\boldsymbol{s}\in\mathcal{S}}d_{\mathrm{TV}}\bigl(P_{S}\bigl(\dot{\boldsymbol{s}}_{t}\mid\dot{\boldsymbol{s}}_{0}=\dot{\boldsymbol{s}}\bigr),\mathbf{P_{\pi_{\boldsymbol{\theta}}}}\bigl(\dot{\boldsymbol{s}}_{t}\bigr)\bigr)\leq\lambda\rho^{t},
\]
for all $t=0,1,\cdots$, where $P_{S}$ denotes the augmented state
distribution, $\mathbf{P_{\pi_{\boldsymbol{\theta}}}}$ is the stationary
case of $P_{S}$ under policy $\pi_{\boldsymbol{\theta}}$, and $d_{\mathrm{TV}}\left(\alpha,\beta\right)=\int_{\dot{\boldsymbol{s}}\in\dot{\mathcal{S}}}\left|\alpha\left(\mathrm{d}\dot{\boldsymbol{s}}\right)-\beta\left(\mathrm{d}\dot{\boldsymbol{s}}\right)\right|$
indicates the total-variation distance between the probability measures
$\alpha$ and $\beta$.\end{assumption}

Assumption 1-1) requires the augmented state space and action spaces
be continuous. Assumption 1-2) can be easily satisfied as long as
there is no gradient explosion during the training process. Assumption
1-3) holds based on the condition that the gradient of DNNs are bounded
and satisfy Lipschitz continuity. Assumption 1-4) controls the bias
caused by the Markovian noise in the observations by assuming the
uniform ergodicity of the Markov chain generated by $\pi_{\boldsymbol{\theta}}$,
which is a standard requirement in \cite{SCAOPO,DQlearning,linnerAC}.

Then, we lay down some assumptions about the choice of algorithm step
sizes:

\begin{assumption}\textit{ (Assumptions on step sizes:)}\\
The step-sizes $\left\{ \eta_{i}\right\} $, $\left\{ \mu_{i}\right\} $,
and $\left\{ \upsilon_{i}\right\} $ are deterministic and non-increasing,
and satisfy:\\
1) $\mu_{i}\rightarrow0$, $\sum_{i}\mu_{i}=\infty$, $\sum_{i}\left(\mu_{i}\right)^{2}<\infty$,\\
2) $\eta_{i}\rightarrow0$, $\frac{1}{\eta_{i}}\leq O\left(i^{\varrho}\right)$
for some $\varrho\in\left(0,1\right)$, $\sum_{i}\left(\eta_{i}\right)^{2}<\infty$,
and $\mathrm{lim}_{i\rightarrow\infty}\mu_{i}\eta_{i}^{-1}=0$,\\
3) $\upsilon_{i}\rightarrow0$, $\sum_{i}\upsilon_{i}=\infty$, $\sum_{i}\left(\upsilon_{i}\right)^{2}<\infty$.\end{assumption}A
typical choice of step sizes step-sizes $\left\{ \eta_{i}\right\} $,
$\left\{ \mu_{i}\right\} $, and $\left\{ \upsilon_{i}\right\} $
satisfying Assumption 2 is $\mu_{i}=O\bigl(i^{-\varrho_{1}}\bigr)$,
$\eta_{i}=O\bigl(i^{-\varrho_{2}}\bigr)$, and $\upsilon_{i}=O\bigl(i^{-\varrho_{3}}\bigr)$,
where $\varrho_{1},\varrho_{2},\varrho_{3}\in\bigl(0.5,1\bigr)$ and
$\varrho_{2}<\varrho_{1}$.

\subsection{Convergence Analyses\label{subsec:The-Asymptotic-Convergence}}

First, we provide the finite-time average estimation error of $\hat{f}$
and $\hat{\boldsymbol{g}}$, which reflects the convergence rate of
surrogate functions in CSSCA method:

\noindent \newtheorem{lemma}{Lemma}\begin{lemma} \textit{(Convergence
rate of the Surrogate Functions:)}\\
The average estimation errors of $\hat{f}$ and $\hat{\boldsymbol{g}}$
are given by\vspace{-0.2cm}
\begin{align*}
\epsilon_{f}\bigl(i\bigr) & \triangleq\frac{1}{1+i}\stackrel[i'=0]{i}{\sum}\mathbb{E}\Bigl[\bigl(\hat{f}_{i'}-\dot{f}\bigl(\boldsymbol{\theta}_{i'}\bigr)\bigr)^{2}\Bigr]\leq O\Bigl(\frac{1}{\bigl(1+i\bigr)\eta_{i+1}}\Bigr)\\
+O\Bigl( & \frac{1}{1+i}\stackrel[i'=0]{i}{\sum}\bigl(\mu_{i'}\mathrm{log}^{2}i+\frac{\mu_{i'}^{2}}{\eta_{i'+1}^{2}}+\frac{\mu_{i'}^{2}}{\eta_{i'+1}}+\eta_{i'+1}\mathrm{log}\thinspace i\bigr)\Bigr),
\end{align*}
\vspace{-0.3cm}
\begin{align*}
\epsilon_{g}\bigl(i\bigr)\triangleq & \frac{1}{1+i}\stackrel[i'=0]{i}{\sum}\mathbb{E}\Bigl[\bigl\Vert\hat{\boldsymbol{g}}_{i'}-\nabla\dot{f}\bigl(\boldsymbol{\theta}_{i'}\bigr)\bigr\Vert_{2}^{2}\Bigr]\leq O\Bigl(\frac{1}{\bigl(1+i\bigr)\eta_{i+1}}\Bigr)\\
+O\Bigl( & \frac{1}{1+i}\stackrel[i'=0]{i}{\sum}\mu_{i'}\mathrm{log}^{2}i+\frac{\mu_{i'}^{2}}{\eta_{i'+1}^{2}}+\frac{\mu_{i'}^{2}}{\eta_{i'+1}}+\eta_{i'+1}\Bigr)\\
+\epsilon_{\mathrm{Q}} & +\sqrt{\epsilon_{f}\bigl(i\bigr)},
\end{align*}
where $\epsilon_{Q}\triangleq O\bigl(\max_{\dot{\boldsymbol{s}},\boldsymbol{a}}\bigl|Q_{\boldsymbol{\omega}_{i}}\bigl(\dot{\boldsymbol{s}},\boldsymbol{a}\bigr)-\hat{Q}^{\pi_{\boldsymbol{\theta}_{i}}}\bigl(\dot{\boldsymbol{s}},\boldsymbol{a}\bigr)\bigr|\bigr)$
represents the error magnitude of the estimated Q-values output Q-networks.
Suppose Assumptions 1 - 2 are satisfied, it is obtained that $\mathrm{lim}_{i\rightarrow\infty}\epsilon_{f}\bigl(i\bigr)=0$
and $\mathrm{lim}_{i\rightarrow\infty}\epsilon_{g}\bigl(i\bigr)=\epsilon_{Q}$.
\end{lemma}
\begin{IEEEproof}
Please refer to Appendix A for details.
\end{IEEEproof}
It has been proven in \cite{DQlearning} that the updates in (\ref{eq:TD-updates})
can reduce $\epsilon_{Q}$ to 0 by increasing the interaction times
$B$ and the update times $T_{\mathrm{cri}}$ in each iteration, as
long as the representational capacity of the Q-networks is sufficiently
strong. Therefore, by choosing appropriate values for $B$, $T_{\mathrm{cri}}$,
and the size of Q-networks, we can ensure that $\epsilon_{Q}$ meets
a practical tolerance standard. More intuitively, if we choose the
typical step sizes me ioned above, use the fact that $\frac{1}{1+i}\stackrel[i'=0]{i}{\sum}\bigl(1+i'\bigr)^{-\varrho}\leq\bigl(1+i'\bigr)^{-\varrho}\bigl(1-\varrho\bigr)^{-1}$,
it can be obtained that
\begin{align*}
\epsilon_{f}\bigl(i\bigr)\leq & O\bigl(i^{\varrho_{1}-1}+i^{-\varrho_{1}}\mathrm{log}i+i^{2\varrho_{1}-2\varrho_{2}}\bigr)\\
\epsilon_{g}\bigl(i\bigr)\leq & O\bigl(i^{\frac{\varrho_{1}-1}{2}}+i^{\frac{-\varrho_{1}}{2}}\mathrm{log}i+i^{\varrho_{1}-\varrho_{2}}+\epsilon_{Q}\bigr).
\end{align*}

Moreover, we also demonstrate that the estimated value $\hat{f}_{k,i}$
and estimated gradient $\hat{\boldsymbol{g}}_{k,i}$ for constructing
surrogate functions satisfy the following asymptotic consistency:\begin{lemma}
\textit{(Asymptotic consistency of surrogate functions)}\\
Suppose that Assumptions 1 and 2 are satisfied, we have
\begin{align}
\underset{t\rightarrow\infty}{\mathrm{lim}} & \bigl|\hat{f}_{k,i}-\dot{f}_{k,i}\left(\boldsymbol{\theta}\right)\bigl|=0,\text{\ensuremath{\forall k,}}\label{eq:JJ}\\
\underset{t\rightarrow\infty}{\mathrm{lim}} & \bigl\Vert\hat{\boldsymbol{g}}_{k,i}-\nabla\dot{f}_{k}\left(\boldsymbol{\theta}_{i}\right)\bigl\Vert_{2}\leq\epsilon_{Q},\text{\ensuremath{\forall k,}}\label{eq:gg}
\end{align}
where $\epsilon_{Q}\triangleq O\bigl(\max_{\dot{\boldsymbol{s}},\boldsymbol{a}}\bigl|Q_{\boldsymbol{\omega}_{i}}\bigl(\dot{\boldsymbol{s}},\boldsymbol{a}\bigr)-\hat{Q}^{\pi_{\boldsymbol{\theta}_{i}}}\bigl(\dot{\boldsymbol{s}},\boldsymbol{a}\bigr)\bigr|\bigr)$.
\end{lemma}
\begin{IEEEproof}
Please refer to Appendix B for details.
\end{IEEEproof}
Then, considering a subsequence $\bigl\{\boldsymbol{\theta}_{i_{j}}\bigr\}_{j=1}^{\infty}$converging
to a limiting point $\boldsymbol{\theta}^{*}$, there exist converged
surrogate functions
\begin{equation}
\underset{j\rightarrow\infty}{\mathrm{lim}}\bar{f}_{k}^{i_{j}}\left(\boldsymbol{\theta}\right)=\hat{f}_{k}\left(\boldsymbol{\theta}\right),\forall\boldsymbol{\theta}\in\mathbf{\Theta},\forall k,\label{eq:converged surrogate function}
\end{equation}
where\vspace{-0.3cm}
\begin{equation}
\bigl|\hat{f}_{k}\left(\boldsymbol{\theta}^{*}\right)-\dot{f}_{k}\left(\boldsymbol{\theta}^{*}\right)\bigl|=0,
\end{equation}
\begin{equation}
\bigl\Vert\nabla\hat{f}_{k}\left(\boldsymbol{\theta}^{*}\right)-\nabla\dot{f}_{k}\left(\boldsymbol{\theta}^{*}\right)\bigl\Vert_{2}=0.
\end{equation}

Finally, with section \ref{subsec:Optimal-Policy-Invariance}, Lemma
1, and Assumptions 1 - 2, we are ready to prove the main convergence
theorem: \noindent \newtheorem{theorem}{Theorem}\begin{theorem}
\textit{(Global Convergence of Algorithm 1:) }\\
Suppose Assumptions 1 -2 are satisfied and the initial point $\boldsymbol{\theta}_{0}$
is feasible, i.e.,$\textrm{max}_{k\in\left\{ 1,\ldots,K\right\} }f_{k}\left(\boldsymbol{\theta}_{0}\right)\leq0$.
Denote $\left\{ \boldsymbol{\theta}_{i}\right\} _{i=1}^{\infty}$
as the iterates generated by the CRL module with a sufficiently small
initial step size $\mu_{0}$. Then, every limiting point $\boldsymbol{\theta}^{*}$
of $\left\{ \boldsymbol{\theta}_{i}\right\} _{i=1}^{\infty}$ satisfying
the Slater condition satisfies KKT conditions up to an error $\epsilon_{Q}$,
i.e.,
\[
\bigl\Vert g_{0}\left(\boldsymbol{\theta}^{*}\right)+\sum_{k}\lambda_{k}g_{k}\left(\boldsymbol{\theta}^{*}\right)\bigl\Vert_{2}\leq\epsilon_{Q}
\]
\[
f_{k}\left(\boldsymbol{\theta}^{*}\right)\leq\epsilon_{Q},k=1,\ldots,K,
\]
\[
\lambda_{k}f_{k}\left(\boldsymbol{\theta}^{*}\right)\leq\epsilon_{Q},k=1,\ldots,K,
\]
where $\epsilon_{Q}\rightarrow0$ as $B\rightarrow\infty$ and $T_{\mathrm{cri}}\rightarrow\infty$
for sufficiently large size of Q-networks.\end{theorem}
\begin{IEEEproof}
The key challenges lie in the proof of Lemma 1 and 2. Once Lemma 1
and 2 are proved, Theorem 1 follows from the similar analyses in our
previous work \cite{Yangrui} and \cite{SCAOPO}, and we omit it due
to the space limit.
\end{IEEEproof}
\begin{rem}
In practice, the pre-trained encoder network and potential networks
unavoidably output imperfect estimates due to the limited number of
updates, the finite representational capacity of DNNs, etc. However,
the simulation results in the next section will demonstrate that these
estimates, despite their errors, can still effectively capture the
current traffic dynamics and packet dropout risks. With these estimates,
the proposed CACRL is able to much better handle non-stationary XR
data traffic with sparse delayed packet loss rewards, compared to
standard CRL algorithms designed for general CMDPs and those incorporating
gradient-based meta-learning methods. Moreover, the CRL module also
contributes to the excellent performance of CACRL, due to its design
for non-convex problems and strict theoretical convergence guarantees.
\end{rem}

\section{Simulation Setup}

We consider a MU-MIMO system with one BS with $M=8$ transmitting
antennas and $K=2\sim12$ single-antenna XR users. We adopt a geometry-based
channel model $\boldsymbol{h}_{k}=\sum_{l=1}^{N_{p}}\bar{\alpha}_{k,l}\mathbf{a}\bigl(\psi_{k,l}\bigr),\forall k$
with $N_{p}$ scattering path following the same setting in \cite{SCAOPO}.
Specifically, the coefficient $\bar{\alpha}_{k,l}\sim\mathcal{CN}\bigl(0,\bar{\sigma}_{k,l}^{2}\bigr)$
is Laplacian distributed, where $\sum_{l=1}^{N_{p}}\bar{\sigma}_{k,l}^{2}=g_{k}$,
$\bar{\sigma}_{k,l}^{2}$ follows an exponential distribution normalized
such that $\sum_{l=1}^{N_{p}}\bar{\sigma}_{k,l}^{2}=g_{k}$, and $g_{k}$
represents the path gain of the $k$-th user. We uniformly generate
the path gains $g_{k}$'s from -10 dB to 10 dB and set $N_{p}=4$
for each user. $\mathbf{a}\bigl(\psi_{k,l}\bigr)$ is the half-wavelength
spaced uniform linear array (ULA) response vector, where $\psi_{k,l}$
denotes the $l$-th angle of departure (AoD) with an angular spread
$\sigma_{AS}=5$. In addition, the bandwidth $W=10$ MHz, and the
noise power density is $-100$ dBm/Hz.

We set the duration of one timeslot to $\tau_{0}=1$ ms. At each timeslot
$t$, we set packets arrive with probabilities $P_{k}$ ranging from
$0.2\sim0.8$. The length $b_{k,t}$ generating from a Poisson distribution
$P_{\mathrm{poi}}\bigl(b_{k}=x\bigr)=\frac{\left(\lambda_{k}\right)^{x}}{x!}e^{-\lambda_{k}}$,
where $\lambda_{k,t}$ randomly ranging from $5\sim20$ Kbits. The
allowed maximum packet dropout rate $c_{k=1,2,\ldots,K}$ are set
to $10\%$. A similar packet arrival model and parameter settings
are also adopted in \cite{XR}, \cite{harddelay4}, and \cite{model0free0RL1}.
All CRL algorithms update their policies every 200 timeslots ($B=200$).
In stationary scenarios, we assume that $P_{k}$ and $\lambda_{k}$
remain unchanged, while in non-stationary scenarios they are regenerated
on average every $E$ timeslots (we refer to $E$ timeslots as one
episode in the following). For the CACRL algorithm, we pre-train the
CI module for 300 episodes, $\zeta_{k}=1,\forall k$, and the step
size parameters are chosen as $\varrho_{1}=0.6,\varrho_{2}=0.7,\varrho_{3}=0.3$.
All simulation results below are averaged across 10 random seeds.

\subsection{Performance Analysis}

\subsubsection{Performance in XR with Stationary Data Traffic}

\begin{figure}[t]
\centering{}\includegraphics[width=6.5cm,height=10cm]{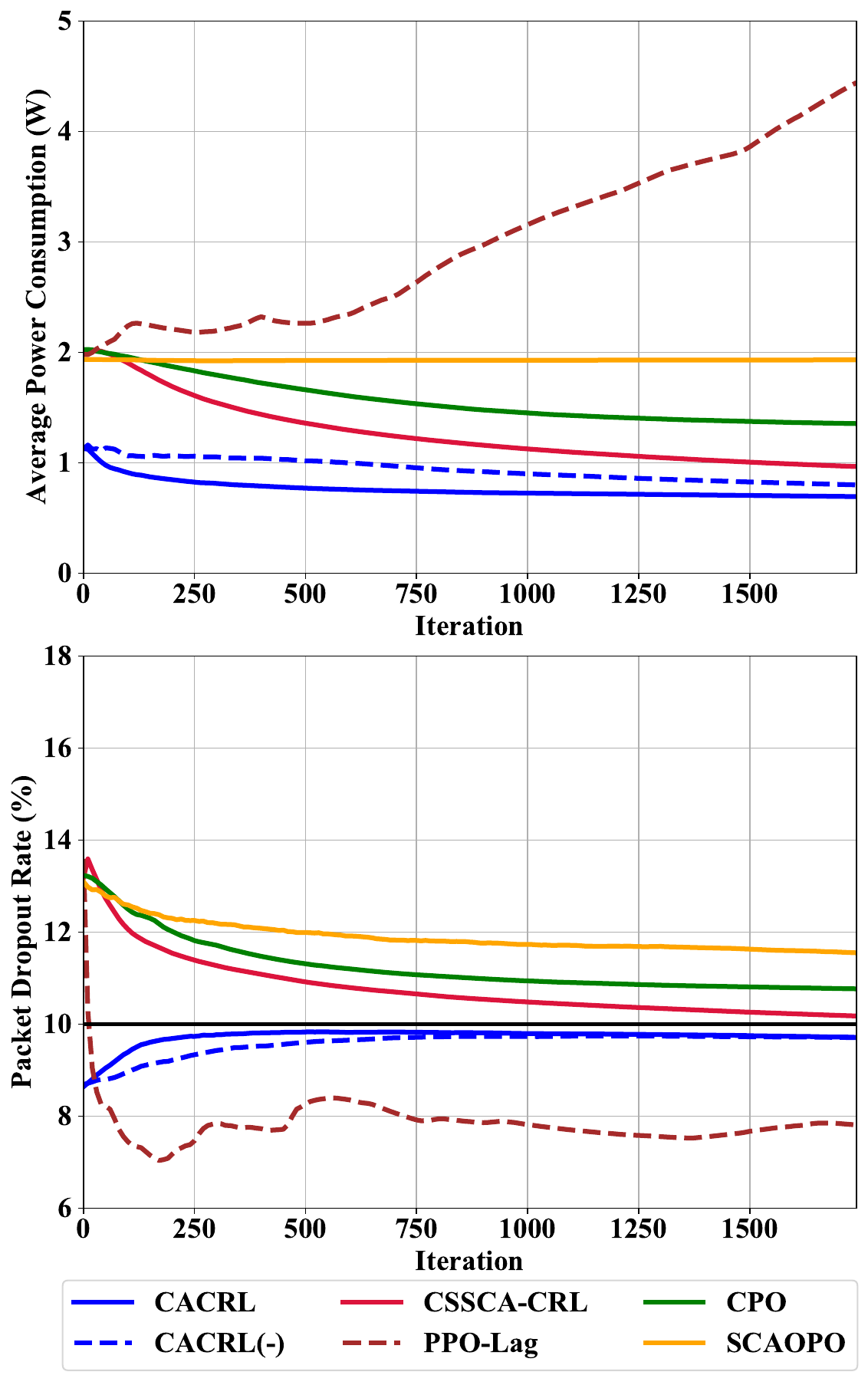}\vspace{-0.2cm}
\caption{\label{fig:stationary performance}Convergence curves in scenario
with stationary XR data traffic.}
\vspace{-0.5cm}
\end{figure}
 We begin with testing algorithms in stationary XR scenarios. The
number of users $K=4$, and the parameters $P_{k}$ and $\lambda_{k}$
are ranging from $\bigl[0.4,0.6\bigr]$ and $\bigl[10,15\bigr]$ Kbits.
Our baseline algorithms include two classical CDAC algorithms, CPO
and PPO-Lag \cite{PPOLagTRPOLag}, and an advanced Actor-only algorithm,
SCAOPO \cite{SCAOPO}, introduced in Section \ref{sec:Introduction}.
All these baselines are standard CRL algorithms designed for stationary
environments, with their policy networks taking only the observable
state $\boldsymbol{s}$ as input. Moreover, we also simulate two variants
of CACRL for ablation experiments: CSSCA-CRL and CACRL(-), where the
former does not employ the CI module, and the latter only incorporates
context-aware meta-learning method.

In Fig. \ref{fig:stationary performance}, it can be seen that since
the policy optimization methods in CPO and PPO-Lag are only suited
for convex constrained problem, they are less effective than CACRL
and its variants in satisfying packet dropout rate constraints and
saving transmission power. Moreover, SCAOPO also struggles to satisfy
the constraints due to the inaccurate Q-value estimation. Comparing
CACRL(-) with CSSCA-CRL, we can see that the former converges faster,
as its pre-trained encoder provides the CRL module with additional
information about the environment dynamics.The results also show that
CACRL converges faster than CACRL(-), highlighting the superiority
of the reward-reshaping method. 
\begin{table*}[t]
\centering{}{\small\caption{\textcolor{blue}{\label{tab:Packet sizes}}Performance versus the
packet sizes}
}{\small{}%
\begin{tabular}{|c|c|c|c|c|c|c|c|c|c|}
\hline 
Algorithm & \multicolumn{3}{c|}{{\small CADAC}} & \multicolumn{3}{c|}{{\small MAML CSSCA-CRL}} & \multicolumn{3}{c|}{{\small MAML CPO}}\tabularnewline
\hline 
Size & {\small Short} & {\small Medium} & {\small Large} & {\small Short} & {\small Medium} & {\small Large} & {\small Short} & {\small Medium} & {\small Large}\tabularnewline
\hline 
{\small Power Consumption (W)} & 0.76 & 0.99 & 1.36 & 0.95 & 1.22 & 2.06 & 1.14 & 1.27 & 2.86\tabularnewline
\hline 
{\small Packet Dropout Rate (\%)} & 9.19 & 9.34 & 10.56 & 9.94 & 12.20 & 12.64 & 12.45 & 13.41 & 18.35\tabularnewline
\hline 
\end{tabular}}\vspace{-0.5cm}
\end{table*}

\subsubsection{Performance in XR with Non-stationary Data Traffic}

\begin{figure}
\centering{}\includegraphics[width=6.5cm,height=10cm]{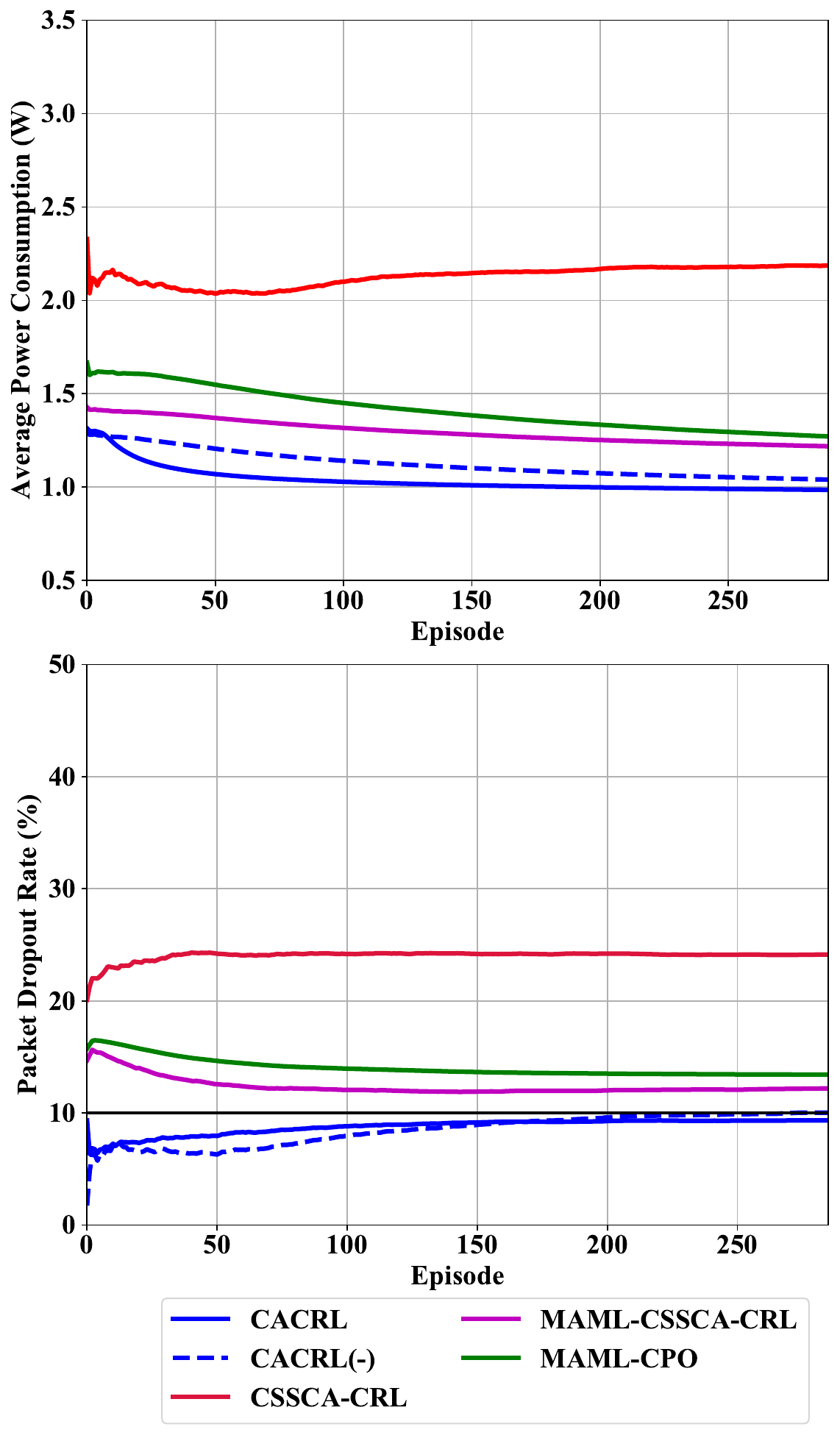}\vspace{-0.4cm}
\caption{\label{fig:non-stationary performance}Convergence curves in scenario
with non-stationary XR data traffic.}
\vspace{-0.6cm}
\end{figure}
Then, we evaluate algorithms in a more challenging XR scenarios with
non-stationary data traffic. The number of users $K=4$, and the parameters
$P_{k}$ and $\lambda_{k}$ are also ranging from $\bigl[0.4,0.6\bigr]$
and $\bigl[10,15\bigr]$ Kbits. Since there are almost no existing
CDAC algorithms for non-stationary environments apart from the MAML-CPO
in \cite{MAML-CPO}, we chose MAML-CPO as a baseline for comparison
with our algorithm. Note that MAML typically belongs to the gradient-based
meta-learning methods introduced in Section \ref{sec:Introduction}.
Additionally, we simulated an algorithm that combines the MAML with
the CRL module of CACRL, namely MAML-CSSCA-CRL. Note that MAML-CPO
and MAML-CSSCA-CRL are also pre-trained for 300 episodes. Moreover,
CACRL(-) and CSSCA-CRL are also simulated for ablation experiments.

\begin{figure}
\begin{centering}
\includegraphics[width=6.5cm,height=6.8cm]{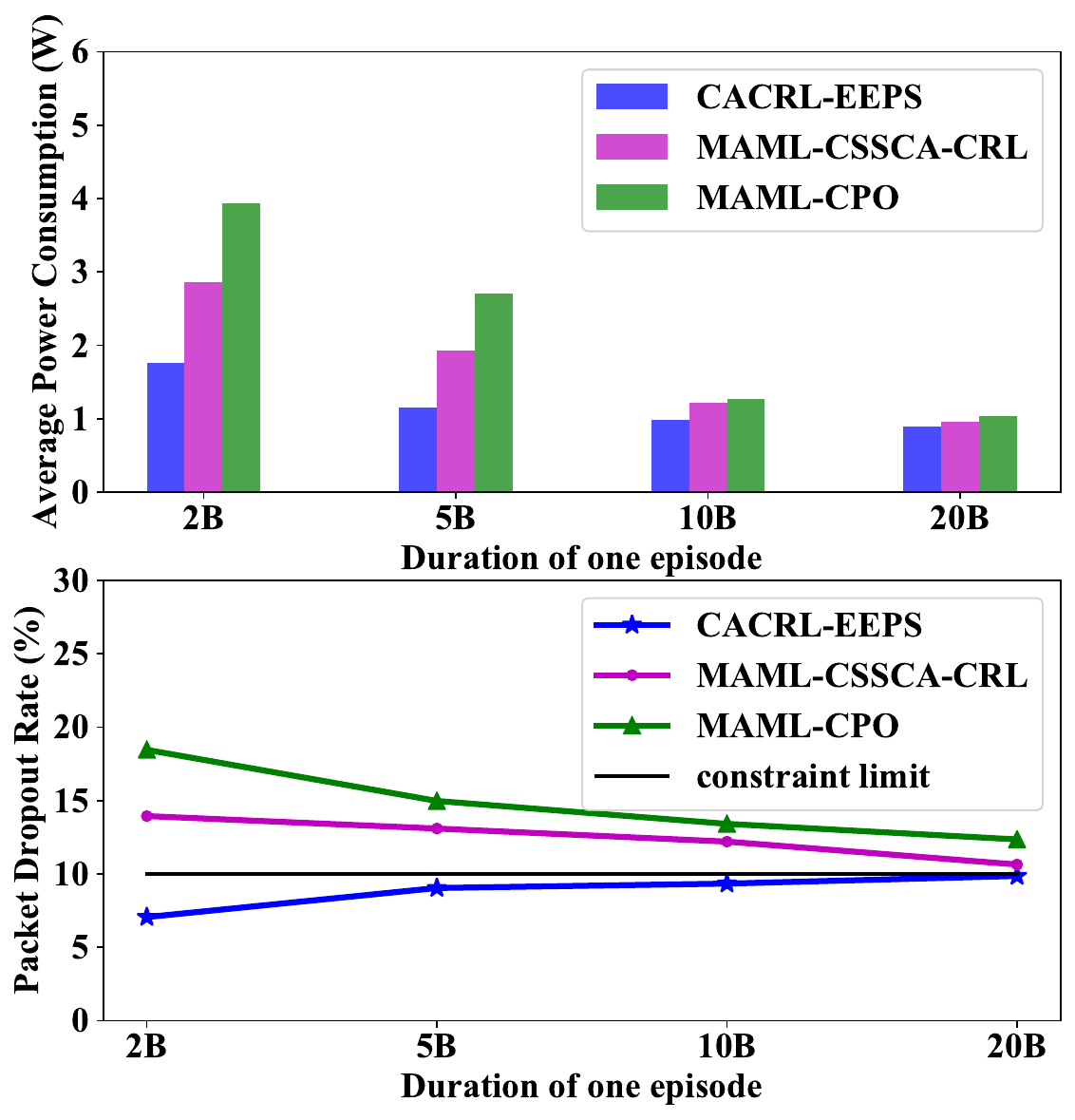}
\par\end{centering}
\centering{}\vspace{-0.3cm}
\caption{\label{fig:varying rates}Performance versus different dynamic varying
rates.}
\vspace{-0.6cm}
\end{figure}

Fig. \ref{fig:non-stationary performance} shows the convergence curves
of the algorithms in a typical scenario with the duration of each
episode $E=10B$ timeslots. In the ablation experiments, the proposed
CACRL outperforms the CACRL(-) both in convergence speed and power
consumption, further demonstrating the effectiveness of reward-reshaping.
Additionally, it is evident that CSSCA-CRL performs poorly, highlighting
the importance of designing CI module for CRL algorithms to handle
the challenges posed by non-stationary XR data traffic. Regarding
the comparative experiments, it is evident that although MAML CSSCA-CDAC
also incorporates the meta-learning method MAML, it does not perform
as well as CACRL and CACRL(-). This is primarily because MAML, as
a gradient-based meta-learning approach, is less effective compared
to context-based meta-learning methods in rapidly changing environments.
Furthermore, it can be seen that the proposed CACRL algorithm greatly
surpasses MAML-CPO due to its superior performance in both the CRL
module and the CI module. For more details, we also tested MAML-CPO
and MAML-CSSCA-CRL alongside the proposed CACRL in scenarios with
the duration of each episode $E=2B,5B,10B,20B$ timeslots, respectively.
Fig. \ref{fig:varying rates} indicates that under rapidly changing
packet arrival dynamics, the MAML-CSSCA-CRL algorithm struggles to
meet constraints and the MAML-CPO almost fails to work. In contrast,
because CACRL is designed to prioritize satisfying constraints with
the feasible update in (\ref{eq:feasible update}) before reducing
power consumption with the objective update in (\ref{eq:objective update}),
it tends to learn a 'safe' policy, resulting in a packet dropout rate
somewhat below the allowable limit. Although this may cause power
consumption to be higher in scenarios with rapidly changing packet
arrival dynamics compared to those with slower dynamics, it remains
lower than that of the other two baselines in the same scenario due
to its superior policy optimization method and advanced CI module.

\subsubsection{Impact of Packet Sizes}

Moreover, we pre-train algorithms in scenarios with different packet
sizes, and show their performance at the deployment stage in Tab.
\ref{tab:Packet sizes}. Specifically, in scenarios with packet sizes
of ``{\small Short}'', ``{\small Medium}'', and ``{\small Large}'',
$P_{k}$ ranges from $\bigl[0.6,0.8\bigr]$, $\bigl[0.4,0.6\bigr]$,
and $\bigl[0.2,0.4\bigr]$, respectively, while $\lambda_{k}$ ranges
from $\bigl[5,10\bigr]$ Kbits, $\bigl[10,15\bigr]$ Kbits, and $\bigl[15,20\bigr]$
Kbits, respectively. Generally, scenarios with long packet sizes are
more critical and challenging because larger packets indicate sparse
feedback rewards, which can easily lead to algorithm convergence failures.
It can be observed that in scenarios with large packets, the MAML-CSSCA-CRL
algorithm requires relatively higher power consumption, and the performance
of MAML-CPO deteriorates sharply to the extent that even with higher
power consumption, it cannot provide the reliability needed to meet
the packet loss constraints. In contrast, our algorithm maintains
almost the same performance across all these scenarios, further demonstrating
the effectiveness of the reward-shaping mechanism.

\subsubsection{Impact of Number of Users}

\begin{figure}
\centering{}\includegraphics[width=6.5cm,height=6.9cm]{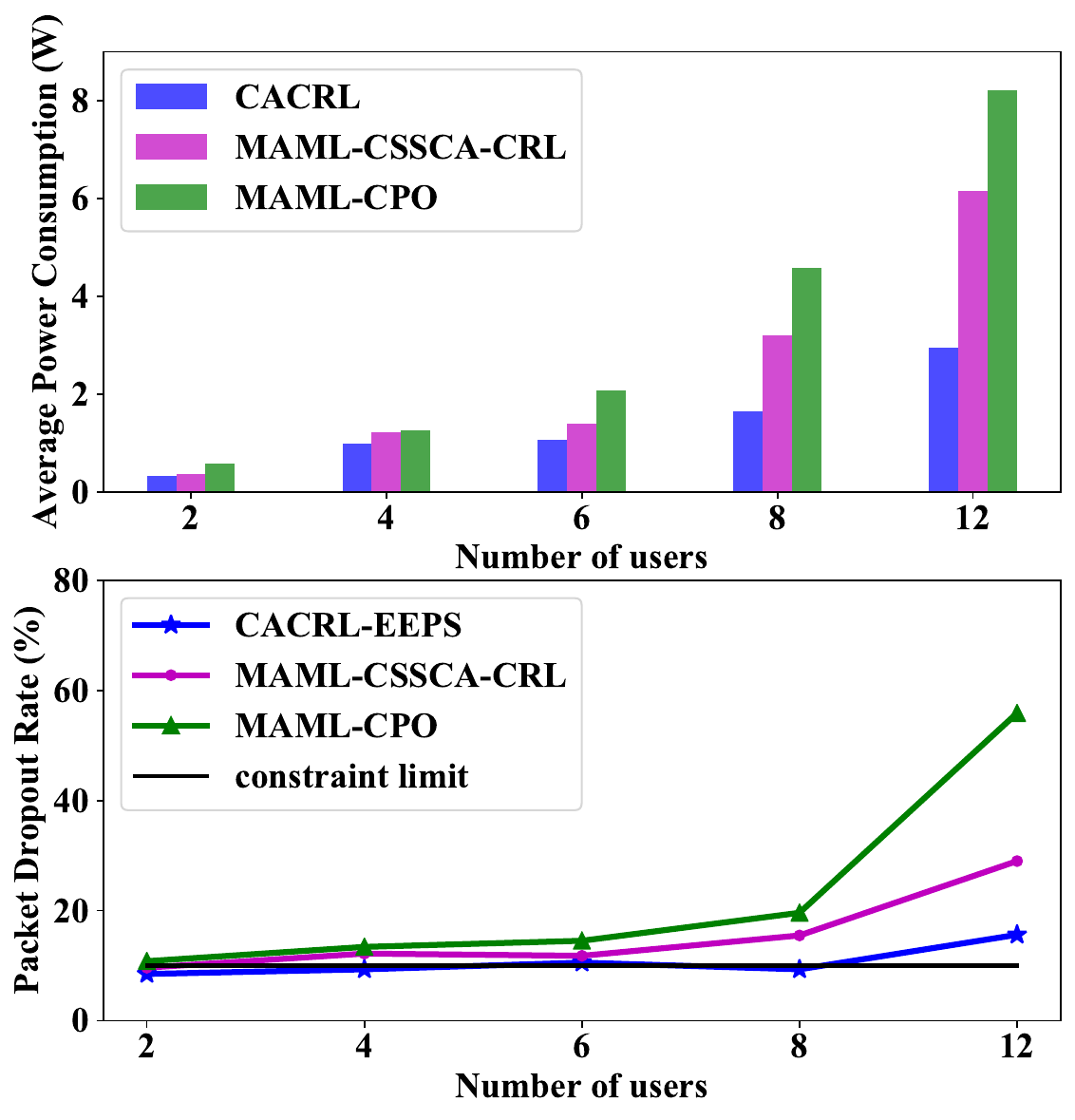}\vspace{-0.3cm}
\caption{\label{fig:number of users}Performance versus the number of users.}
\vspace{-0.5cm}
\end{figure}

Finally, we also pre-train algorithms in scenarios with different
numbers of users, where the parameters $P_{k}$ and $\lambda_{k}$
are ranging from $\bigl[0.4,0.6\bigr]$ and $\bigl[10,15\bigr]$ Kbits.
Fig. \ref{fig:number of users} presents performance of algorithms
in scenarios with different numbers of users. First, it can be seen
that our algorithm uses less power resources compared to the other
two baselines across various scenarios. Then, when it comes to guaranteeing
QoS constraints, all algorithms perform better with fewer users but
deteriorate as the number of users increases. This is because with
more users, the EEPS problem becomes more non-convex and more complex
packet arrival processes, leading to worse performance and slower
convergence of algorithms. However, due to the the CSSCA-based policy
optimization method, which is better suited for non-convex problems,
and the advanced CI module, our algorithm still outperforms the other
two baseline algorithms even with a larger number of users.

\section{Conclusion}

This paper investigates a novel CACRL algorithm composed of a CRL
module and a CI module, aiming at supporting XR transmission with
minimal transmitted power resources. The CACRL formulates the EEPS
task as a DP-CMDP, which appropriately capture its requirements and
characteristics. In the CI module, we adopt the context-aware meta-learning
method and the reward-reshaping method to transform the original DP-CMDP
to a general CMDP with immediate dense packet dropout rewards, enabling
the CACRL algorithm to effectively handle non-stationary XR data traffic,
even when packet dropout rewards are sparse and delayed. In the CRL
module, we introduce a CSSCA-based policy optimization method, which
can address non-convex stochastic constraints more effectively than
conventional methods. Theoretical analysis is provided to offer deep
insight into the algorithm's convergence, and simulation results demonstrate
that the proposed CACRL algorithm outperforms advanced baselines,
making it highly promising for future XR transmission applications.

\begin{appendices}

\section{Proof of Lemma 1}

Since the analysis for $\epsilon_{f}\bigl(i\bigr)$ and $\epsilon_{g}\bigl(i\bigr)$
in Theorem 1 are similar, we only present the more intractable derivation
of $\epsilon_{g}\bigl(i\bigr)$ due to the space limit. We first define
\[
\boldsymbol{b}_{i}=\hat{\boldsymbol{g}}_{i}-\nabla\dot{f}\bigl(\boldsymbol{\theta}_{i}\bigr),
\]
\[
\varUpsilon\bigl(\varepsilon_{i+1},\hat{\boldsymbol{g}}_{i},\boldsymbol{\theta}_{i},Q_{\boldsymbol{\omega}_{i+1}}\bigr)=\bigl\langle\boldsymbol{b}_{i},\tilde{\boldsymbol{g}}_{i+1}-\nabla\dot{f}\bigl(\boldsymbol{\theta}_{i}\bigr)\bigr\rangle,
\]
where the subscript $k$ is omitted to simplify notation. Then, according
to (\ref{eq:g-hat}), it can be derived that \vspace{-0.2cm}
\begin{align*}
\Bigr\Vert & \boldsymbol{b}_{i+1}\Bigr\Vert_{2}^{2}=\Bigr\Vert\boldsymbol{b}_{i}+\nabla\dot{f}\bigl(\boldsymbol{\theta}_{i}\bigr)-\nabla\dot{f}\bigl(\boldsymbol{\theta}_{i+1}\bigr)+\eta_{i+1}\bigl(\tilde{\boldsymbol{g}}_{i+1}-\hat{\boldsymbol{g}}_{i}\bigr)\Bigr\Vert_{2}^{2}\\
\leq & \Bigr\Vert\boldsymbol{b}_{i}\Bigr\Vert_{2}^{2}+2\eta_{i+1}\boldsymbol{b}_{i}^{\intercal}\Bigl(\tilde{\boldsymbol{g}}_{i+1}-\hat{\boldsymbol{g}}_{i}\Bigr)+2\Bigr\Vert\nabla\dot{f}\bigl(\boldsymbol{\theta}_{i}\bigr)-\nabla\dot{f}\bigl(\boldsymbol{\theta}_{i+1}\bigr)\Bigr\Vert_{2}^{2}\\
 & +2\boldsymbol{b}_{i}^{\intercal}\Bigl(\nabla\dot{f}\bigl(\boldsymbol{\theta}_{i}\bigr)-\nabla\dot{f}\bigl(\boldsymbol{\theta}_{i+1}\bigr)\Bigr)+2\eta_{i+1}^{2}\Bigr\Vert\tilde{\boldsymbol{g}}_{i+1}-\hat{\boldsymbol{g}}_{i}\Bigr\Vert_{2}^{2}\\
= & \bigl(1-2\eta_{i+1}\bigr)\Bigr\Vert\boldsymbol{b}_{i}\Bigr\Vert_{2}^{2}+2\eta_{i+1}\varUpsilon\Bigl(\varepsilon_{i+1},\hat{\boldsymbol{g}}_{i},\boldsymbol{\theta}_{i},Q_{\boldsymbol{\omega}_{i+1}}\Bigr)\\
 & +2\boldsymbol{b}_{i}^{\intercal}\Bigl(\nabla\dot{f}\bigl(\boldsymbol{\theta}_{i}\bigr)-\nabla\dot{f}\bigl(\boldsymbol{\theta}_{i+1}\bigr)\Bigr)+2\Bigr\Vert\nabla\dot{f}\bigl(\boldsymbol{\theta}_{i}\bigr)-\nabla\dot{f}\bigl(\boldsymbol{\theta}_{i+1}\bigr)\Bigr\Vert_{2}^{2}\\
 & +2\eta_{i+1}^{2}\Bigr\Vert\tilde{\boldsymbol{g}}_{i+1}-\hat{\boldsymbol{g}}_{i}\Bigr\Vert_{2}^{2}.
\end{align*}
Rearranging, taking expectation, and taking summation on both sides,
we obtain that\vspace{-0.2cm}
\begin{align}
 & \frac{1}{1+i}\stackrel[i'=0]{i}{\sum}\mathbb{E}\bigl[\bigl\Vert\boldsymbol{b}_{i'}\bigr\Vert_{2}^{2}\bigr]\label{eq:average_error_g}\\
\leq & \underset{A_{1}\left(i\right)}{\underbrace{\frac{1}{1+i}\stackrel[i'=0]{i}{\sum}\frac{1}{2\eta_{i'+1}}\mathbb{E}\Bigl[\bigl\Vert\boldsymbol{b}_{i'}\bigr\Vert_{2}^{2}-\bigl\Vert\boldsymbol{b}_{i'+1}\bigr\Vert_{2}^{2}\bigr]}}\nonumber \\
+ & \underset{A_{2}\left(i\right)}{\underbrace{\frac{1}{1+i}\stackrel[i'=0]{i}{\sum}\frac{1}{\eta_{i'+1}}\mathbb{E}\Bigl[\bigl\Vert\nabla\dot{f}\bigl(\boldsymbol{\theta}_{i'}\bigr)-\nabla\dot{f}\bigl(\boldsymbol{\theta}_{i'+1}\bigr)\bigr\Vert_{2}^{2}}\bigr]}\nonumber \\
+ & \underset{A_{3}\left(i\right)}{\underbrace{\frac{1}{1+i}\stackrel[i'=0]{i}{\sum}\eta_{i'+1}\mathbb{E}\bigl[\bigl\Vert\tilde{\boldsymbol{g}}_{i'+1}-\hat{\boldsymbol{g}}_{i'}\bigr\Vert_{2}^{2}\bigr]}}\Bigr)\nonumber \\
+ & \underset{A_{4}\left(i\right)}{\underbrace{\frac{1}{1+i}\stackrel[i'=0]{i}{\sum}\mathbb{E}\bigl[\varUpsilon\bigl(\varepsilon_{i'+1},\hat{\boldsymbol{g}}_{i'},\boldsymbol{\theta}_{i'},Q_{\boldsymbol{\omega}_{i'+1}}\bigr)\bigr]}}\nonumber \\
+ & \underset{A_{5}\left(i\right)}{\underbrace{\frac{1}{1+i}\stackrel[i'=0]{i}{\sum}\frac{1}{\eta_{i'+1}}\mathbb{E}\bigl[\bigl\langle\boldsymbol{b}_{i'},\nabla\dot{f}\bigl(\boldsymbol{\theta}_{i'}\bigr)-\nabla\dot{f}\bigl(\boldsymbol{\theta}_{i'+1}\bigr)\bigr\rangle\bigr]}},\nonumber 
\end{align}

For the first terem $A_{1}\bigl(i\bigr)$,
\begin{align}
A_{1}\bigl(i\bigr)= & \frac{1}{1+i}\stackrel[i'=0]{i}{\sum}\frac{1}{2\eta_{i'+1}}\mathbb{E}\Bigl[\bigl\Vert\boldsymbol{y}_{i'}\bigr\Vert_{2}^{2}-\bigl\Vert\boldsymbol{y}_{i'+1}\bigr\Vert_{2}^{2}\Bigr]\nonumber \\
= & \frac{1}{1+i}\Bigl(\stackrel[i'=0]{i}{\sum}\bigl(\frac{1}{2\eta_{i'+1}}-\frac{1}{2\eta_{i'}}\bigr)\mathbb{E}\bigl[\bigl\Vert\boldsymbol{b}_{i'}\bigr\Vert_{2}^{2}\bigr]\nonumber \\
 & +\frac{1}{2\eta_{1}}\mathbb{E}\Bigl[\bigl\Vert\boldsymbol{b}_{0}\bigr\Vert_{2}^{2}\Bigr]-\frac{1}{2\eta_{i+1}}\mathbb{E}\bigl[\bigl\Vert\boldsymbol{b}_{i+1}\bigr\Vert_{2}^{2}\bigr]\Bigr)\nonumber \\
\leq & \frac{1}{1+i}\Bigl(\stackrel[i'=0]{i}{\sum}\bigl(\frac{1}{2\eta_{i'+1}}-\frac{1}{2\eta_{i'}}\bigr)b_{\mathrm{max}}\nonumber \\
 & +\frac{1}{2\eta_{1}}b_{\mathrm{max}}-\frac{1}{2\eta_{i'+1}}b_{\mathrm{min}}\Bigr)\nonumber \\
= & O\bigl(\frac{1}{1+i}\frac{1}{\eta_{i+1}}\bigr),\label{eq:M1}
\end{align}
where $b_{\mathrm{max}}$ and $b_{\mathrm{min}}$ are the maximum
and minimum values of $\mathbb{E}\bigl[\left\Vert \boldsymbol{b}_{i'}\right\Vert _{2}^{2}\bigr],\forall i'$,
respectively.

In terms of $A_{2}\bigl(i\bigr)$ and $A_{3}\bigl(i\bigr)$, we respectively
have that\vspace{-0.2cm}
\begin{align}
A_{2}\bigl(i\bigr)= & \frac{1}{1+i}\stackrel[i'=0]{i}{\sum}\frac{1}{\eta_{i'+1}}\mathbb{E}\Bigl[\bigl\Vert\nabla\dot{f}\bigl(\boldsymbol{\theta}_{i'}\bigr)-\nabla\dot{f}\bigl(\boldsymbol{\theta}_{i'+1}\bigr)\bigr\Vert_{2}^{2}\bigr]\nonumber \\
\overset{a}{=} & O\bigl(\frac{1}{1+i}\stackrel[i'=0]{i}{\sum}\frac{\mu_{i'}^{2}}{\eta_{i'+1}}\bigr),\label{eq:A2}
\end{align}
\vspace{-0.3cm}
\begin{align}
A_{3}\bigl(i\bigr)= & \frac{1}{1+i}\stackrel[i'=0]{i}{\sum}\eta_{i'+1}\mathbb{E}\bigl[\bigl\Vert\tilde{\boldsymbol{g}}_{i'+1}-\hat{\boldsymbol{g}}_{i'}\bigr\Vert_{2}^{2}\bigr]\label{eq:A3}\\
\overset{a}{=} & O\bigl(\frac{1}{1+i}\stackrel[i'=0]{i}{\sum}\eta_{i'+1}\bigr),\nonumber 
\end{align}
where (\ref{eq:A2})-a follows from the Lipschitz continuity of $\nabla\dot{f}\bigl(\boldsymbol{\theta}\bigr)$,
and (\ref{eq:A3})-a utilize the boundedness of $\bigl\Vert\tilde{\boldsymbol{g}}_{i+1}-\hat{\boldsymbol{g}}_{i}\bigr\Vert_{2}^{2}$.

For $A_{4}\bigl(i\bigr)$, we denote a gradient term
\begin{equation}
\nabla\hat{f}\left(\boldsymbol{\theta}\right)=\mathbb{E}\left[\hat{Q}^{\pi_{\boldsymbol{\theta}}}\left(\dot{\boldsymbol{s}},\boldsymbol{a}\right)\nabla_{\boldsymbol{\theta}}\textrm{log}\pi_{\boldsymbol{\theta}}\left(\boldsymbol{a}\mid\dot{\boldsymbol{s}}\right)\right],\label{eq:auxiliary policy gradient-1}
\end{equation}
and further denote $\varUpsilon\bigl(\varepsilon_{i+1},\hat{\boldsymbol{g}}_{i},\boldsymbol{\theta}_{i},\hat{Q}^{\pi_{\boldsymbol{\theta}_{i}}}\bigr)=\bigl\langle\boldsymbol{b}_{i},\nabla\hat{f}\bigl(\boldsymbol{\theta}_{i}\bigr)-\nabla f\bigl(\boldsymbol{\theta}_{i}\bigr)\bigr\rangle$
and $\varUpsilon\bigl(\hat{\boldsymbol{g}}_{i},\boldsymbol{\theta}_{i}\bigr)=\bigl\langle\boldsymbol{b}_{i},\nabla\dot{f}\bigl(\boldsymbol{\theta}_{i}\bigr)-\nabla\dot{f}\bigl(\boldsymbol{\theta}_{i}\bigr)\bigr\rangle=0$.
Then, we derive that
\begin{align}
 & \mathbb{E}\bigl[\varUpsilon\bigl(\varepsilon_{i'+1},\hat{\boldsymbol{g}}_{i'},\boldsymbol{\theta}_{i'},Q_{\boldsymbol{\omega}_{i'+1}}\bigr)\bigr]\nonumber \\
= & \mathbb{E}\bigl[\varUpsilon\bigl(\varepsilon_{i'+1},\hat{\boldsymbol{g}}_{i'},\boldsymbol{\theta}_{i'},Q_{\boldsymbol{\omega}_{i'+1}}\bigr)-\varUpsilon\bigl(\varepsilon_{i'+1},\hat{\boldsymbol{g}}_{i'},\boldsymbol{\theta}_{i'},\hat{Q}^{\pi_{\boldsymbol{\theta}_{i'+1}}}\bigr)\bigr]\nonumber \\
 & +\mathbb{E}\bigl[\varUpsilon\bigl(\varepsilon_{i'+1},\hat{\boldsymbol{g}}_{i'},\boldsymbol{\theta}_{i'},\hat{Q}^{\pi_{\boldsymbol{\theta}_{i'+1}}}\bigr)-\varUpsilon\bigl(\hat{\boldsymbol{g}}_{i'},\boldsymbol{\theta}_{i'}\bigr)\bigr]\nonumber \\
\overset{a}{\leq} & \Bigl|\mathbb{E}\bigl[\varUpsilon\bigl(\varepsilon_{i'+1},\hat{\boldsymbol{g}}_{i'},\boldsymbol{\theta}_{i'},Q_{\boldsymbol{\omega}_{i'+1}}\bigr)-\varUpsilon\bigl(\varepsilon_{i'+1},\hat{\boldsymbol{g}}_{i'},\boldsymbol{\theta}_{i'},\hat{Q}^{\pi_{\boldsymbol{\theta}_{i'+1}}}\bigr)\bigr]\Bigr|\nonumber \\
 & +\Bigl|\mathbb{E}\bigl[\varUpsilon\bigl(\varepsilon_{i'+1},\hat{\boldsymbol{g}}_{i'},\boldsymbol{\theta}_{i'},\hat{Q}^{\pi_{\boldsymbol{\theta}_{i'+1}}}\bigr)-\varUpsilon\bigl(\hat{\boldsymbol{g}}_{i'},\boldsymbol{\theta}_{i'}\bigr)\bigr]\Bigr|.\label{eq:R}
\end{align}
Denoting the first term of (\ref{eq:R})-a as $D_{1}\bigl(i'\bigr)$,
we have
\begin{align}
D_{1}\bigl(i'\bigr)= & O\bigl(\mathbb{E}\bigl[\bigl\Vert\tilde{\boldsymbol{g}}_{i'+1}-\nabla\hat{f}\bigl(\boldsymbol{\theta}_{i'+1}\bigr)\bigr\Vert_{2}\bigr]\bigr)\label{eq:D1}\\
\overset{a}{\leq} & \frac{1}{B}\stackrel[j=Bi'+1]{Bi'+B}{\sum}O\bigl(\bigl\Vert P_{S}\bigl(\dot{\boldsymbol{s}}_{j}\bigr)-\mathbf{P}_{\pi_{\boldsymbol{\theta}_{i'+1}}}\bigl(\dot{\boldsymbol{s}}_{j}\bigr)\bigr\Vert_{\mathrm{TV}}\nonumber \\
+ & \bigl|\mathbb{E}\bigl[Q_{\boldsymbol{\omega}_{i'+1}}\left(\dot{\boldsymbol{s}},\boldsymbol{a}\right)\bigr]-\hat{Q}^{\pi_{\boldsymbol{\theta}_{i'}+1}}\left(\dot{\boldsymbol{s}},\boldsymbol{a}\right)\bigr|\bigr)\nonumber 
\end{align}
where (\ref{eq:D1})-a is derived according to the definations of
$\tilde{\boldsymbol{g}}_{i'+1}$ and $\nabla\hat{f}\bigl(\boldsymbol{\theta}_{i'+1}\bigr)$
and the triangle inequality. By denoting an auxiliary trajectory $\bigl\{\dot{\tilde{\boldsymbol{s}}}_{j-\kappa_{i}},\tilde{\boldsymbol{a}}_{j-\kappa_{i}},\dot{\tilde{\boldsymbol{s}}}_{j-\kappa_{i}+1},\tilde{\boldsymbol{a}}_{j-\kappa_{i}+1},\ldots,\dot{\tilde{\boldsymbol{s}}}_{j}\bigr\}$
obtained by the fixed policy $\pi_{\boldsymbol{\theta}_{i'+1}}$ from
the initial state $\dot{\tilde{\boldsymbol{s}}}_{j-\kappa_{i}}=\dot{\boldsymbol{s}}_{j-\kappa_{i}}$,
the first term of (\ref{eq:D1}) can be derived as 
\begin{align}
 & O\bigl(\bigl\Vert P_{S}\bigl(\dot{\boldsymbol{s}}_{j}\bigr)-\mathbf{P}_{\pi_{\boldsymbol{\theta}_{i'+1}}}\bigl(\dot{\boldsymbol{s}}_{j}\bigr)\bigr\Vert_{\mathrm{TV}}\label{eq:TV-P}\\
\overset{a}{\leq} & O\bigl(\bigl\Vert P_{S}\bigl(\dot{\boldsymbol{s}}_{j}\bigr)-P_{S}\bigl(\dot{\tilde{\boldsymbol{s}}}_{j-\kappa_{i}}\bigr)\bigr\Vert_{\mathrm{TV}}\nonumber \\
 & +\bigl\Vert P_{S}\bigl(\dot{\tilde{\boldsymbol{s}}}_{j-\kappa_{i}}\bigr)-\mathbf{P}_{\pi_{\boldsymbol{\theta}_{i'+1}}}\bigl(\dot{\boldsymbol{s}}_{j}\bigr)\bigr\Vert_{\mathrm{TV}}\bigr).\nonumber 
\end{align}
Following the similar tricks as (23)-(26) in \cite{SCAOPO}, we have
\begin{align}
\bigl\Vert P_{S}\bigl(\dot{\boldsymbol{s}}_{j}\bigr)-P_{S}\bigl(\dot{\tilde{\boldsymbol{s}}}_{j-\kappa_{i}}\bigr)\bigr\Vert_{\mathrm{TV}} & \leq O\bigl(\kappa_{i}\sum_{l=j-\kappa_{i}}^{j}\mu_{\left\lfloor l/B\right\rfloor }\bigr).\label{eq:term1}
\end{align}
According to Assumption 1-4), it is obtained that
\begin{equation}
\bigl\Vert P_{S}\bigl(\dot{\tilde{\boldsymbol{s}}}_{j-\kappa_{i}}\bigr)-\mathbf{P}_{\pi_{\boldsymbol{\theta}_{i'+1}}}\bigl(\dot{\boldsymbol{s}}_{j}\bigr)\bigr\Vert_{\mathrm{TV}}=O\bigl(\rho^{\kappa_{i}}\bigr).\label{eq:term2}
\end{equation}
 Denoting the second term of (\ref{eq:D1})-a as $D_{2}\bigl(i\bigr)$,
we have
\begin{align}
D_{2}\bigl(i'\bigr) & =\mathbb{E}\Bigl[\bigl\Vert\nabla\hat{f}\bigl(\boldsymbol{\theta}_{i'}+1\bigr)-\nabla\dot{f}\bigl(\boldsymbol{\theta}_{i'}\bigr)\bigr\Vert_{2}\Bigr]\nonumber \\
 & =O\bigl(\mathbb{E}\bigl[\bigl|\hat{f}_{i'}-\dot{f}\bigl(\boldsymbol{\theta}_{i'}\bigr)\bigr|\bigr]\bigr).\label{eq:D2}
\end{align}
Therefore, combining (\ref{eq:D1})-(\ref{eq:D2}) and setting $O\left(\rho^{\kappa_{i}}\right)=O\left(\frac{1}{i}\right)$,
i.e., $\kappa_{i}=O\left(\mathrm{log}\thinspace i\right)$, we can
obtain that
\begin{align}
A_{4}\bigl(i\bigr) & =O\Bigl(\frac{1}{1+i}\stackrel[i'=0]{i}{\sum}\mathbb{E}\Bigl[\bigl|\hat{f}_{i'}-\dot{f}\bigl(\boldsymbol{\theta}_{i'}\bigr)\bigl|\Bigr]\label{eq:M4}\\
 & +\frac{1}{1+i}\stackrel[i'=0]{i}{\sum}\bigl(\mu_{i'}\mathrm{log}^{2}i+\frac{1}{i}+\epsilon_{\mathrm{cri}}\bigl(i'\bigr)\Bigr)\Bigr)\nonumber \\
 & \overset{a}{\leq}O\Bigl(\sqrt{\frac{1}{1+i}\stackrel[i'=0]{i}{\sum}\mathbb{E}\Bigl[\bigl|\hat{f}_{i'}-\dot{f}\bigl(\boldsymbol{\theta}_{i'}\bigr)\bigl|^{2}\Bigr]}\nonumber \\
 & +\frac{1}{1+i}\stackrel[i'=0]{i}{\sum}\bigl(\mu_{i'}\mathrm{log}^{2}i+\frac{1}{i}+\epsilon_{Q}\bigl(i'\bigr)\bigr)\Bigr)\nonumber 
\end{align}

For $A_{5}\bigl(t\bigr)$, we have{\small
\begin{align}
 & A_{5}\bigl(i\bigr)=\frac{1}{1+i}\stackrel[i'=0]{i}{\sum}\frac{1}{\eta_{i'+1}}\mathbb{E}\Bigl[\boldsymbol{b}_{i'}^{\intercal}\Bigl(\nabla\dot{f}\bigl(\boldsymbol{\theta}_{i'}\bigr)-\nabla\dot{f}\bigl(\boldsymbol{\theta}_{i'+1}\bigr)\Bigr)\Bigr]\nonumber \\
 & \overset{a}{\leq}\sqrt{\frac{1}{1+i}\stackrel[i'=0]{i}{\sum}\mathbb{E}\Bigl[\bigl\Vert\boldsymbol{b}_{i'}\bigr\Vert_{2}^{2}\Bigr]}\sqrt{\frac{1}{1+i}\stackrel[i'=0]{i}{\sum}\frac{\bigl(\nabla\dot{f}\bigl(\boldsymbol{\theta}_{i'}\bigr)-\nabla\dot{f}\bigl(\boldsymbol{\theta}_{i'+1}\bigr)\bigr)^{2}}{\eta_{i'+1}^{2}}}\nonumber \\
 & \overset{b}{=}O\Bigl(\sqrt{B\bigl(i\bigr)}\sqrt{\frac{1}{1+i}\stackrel[i'=0]{i}{\sum}\frac{\mu_{i'}^{2}}{\eta_{i'+1}^{2}}}\Bigr),\label{eq:M5}
\end{align}
}where we define $B\bigl(i\bigr)\overset{\Delta}{=}\frac{1}{1+i}\stackrel[i'=0]{i}{\sum}\mathbb{E}\bigl[\bigl\Vert\boldsymbol{b}_{i'}\bigr\Vert_{2}^{2}\bigr]$.
Please note that (\ref{eq:M5})-a follows from the Cauchy-Schwarz
inequality, and (\ref{eq:M5})-b is because of the Lipschitz continuity
of $\nabla\dot{f}\bigl(\boldsymbol{\theta}\bigr)$ and the update
rule $\boldsymbol{\theta}_{i'}=\boldsymbol{\theta}_{i'-1}+\eta_{i'}\bigl(\boldsymbol{\bar{\theta}}_{i'}-\boldsymbol{\theta}_{i'-1}\bigr)$.

Combining equations (\ref{eq:M1})-(\ref{eq:M5}) and defining
\[
C\bigl(i\bigr)=\frac{1}{1+i}\stackrel[i'=0]{i}{\sum}\frac{\mu_{i'}^{2}}{\eta_{i'+1}^{2}},
\]
\[
A\bigl(i\bigr)=A_{1}\bigl(i\bigr)+A_{2}\bigl(i\bigr)+A_{3}\bigl(i\bigr)+A_{4}\bigl(i\bigr).
\]
we have
\begin{align*}
B\bigl(i\bigr)= & O\Bigl(\sqrt{B\bigl(i\bigr)}\sqrt{C\bigl(i\bigr)}+A\bigl(i\bigr)\Bigr)\\
\leq & C_{1}\sqrt{B\bigl(i\bigr)}\sqrt{C\bigl(i\bigr)}+O\Bigl(A\bigl(i\bigr)\Bigr),
\end{align*}
where $c_{0}>0$ is some positive constant. Then, it is obtained 
\begin{align}
\Bigl(\sqrt{B\bigl(i\bigr)}-\frac{c_{0}}{2}\sqrt{C\bigl(i\bigr)}\Bigr)^{2} & \leq\frac{c_{0}^{2}}{4}C\bigl(i\bigr)+O\Bigl(A\bigl(i\bigr)\Bigr)\label{eq:Ft_equation1}\\
\sqrt{B\bigl(i\bigr)}-\frac{c_{0}}{2}\sqrt{C\bigl(i\bigr)} & \leq\frac{c_{0}}{2}\sqrt{C\bigl(i\bigr)}+O\Bigl(A\bigl(i\bigr)\Bigr)\nonumber \\
\sqrt{B\bigl(i\bigr)} & \leq c_{0}\sqrt{C\bigl(i\bigr)}+O\Bigl(A\bigl(i\bigr)\Bigr)\nonumber \\
B\bigl(i\bigr) & =O\Bigl(C\bigl(i\bigr)+A\bigl(i\bigr)\Bigr).\nonumber 
\end{align}
 This completes the proof.

\section{Proof of Lemma 2}

Since the asymptotic consistency analyses for function values and
gradients are similar, we will only present the latter.We first construct
two auxiliary policy gradient estimates $\nabla_{\boldsymbol{\theta}}\hat{f}_{k}\left(\boldsymbol{\theta}_{i}\right)$
and $\hat{\bar{\boldsymbol{g}}}_{k,i}$. Specifically, $\nabla_{\boldsymbol{\theta}}\hat{f}_{k}\left(\boldsymbol{\theta}\right)$
is defined by
\begin{equation}
\nabla_{\boldsymbol{\theta}}\hat{f}_{k}\left(\boldsymbol{\theta}_{i}\right)=\mathbb{E}_{\sigma_{\pi_{\boldsymbol{\theta}_{i}}}}\left[\hat{Q}_{k}^{\pi_{\boldsymbol{\theta}_{i}}}\left(\dot{\boldsymbol{s}},\boldsymbol{a}\right)\nabla_{\boldsymbol{\theta}}\textrm{log}\pi_{\boldsymbol{\theta}_{i}}\left(\boldsymbol{a}\mid\dot{\boldsymbol{s}}\right)\right],\forall k.\label{eq:auxiliary policy gradient}
\end{equation}
and $\hat{\bar{\boldsymbol{g}}}_{k,i}$ is obtained by\vspace{-0.3cm}
\begin{equation}
\hat{\bar{\boldsymbol{g}}}_{k,i+1}=\left(1-\eta_{i+1}\right)\hat{\bar{\boldsymbol{g}}}_{k,i}+\eta_{i}\tilde{\bar{\boldsymbol{g}}}_{k,i},\forall k,i,
\end{equation}
\begin{equation}
\tilde{\bar{\boldsymbol{g}}}_{k,i}=\hat{\mathbb{E}}_{\varepsilon_{i}}\left[\hat{Q}_{k}^{\pi_{\boldsymbol{\theta}_{i}}}\left(\dot{\boldsymbol{s}},\boldsymbol{a}\right)\nabla_{\boldsymbol{\theta}}\textrm{log}\pi_{\boldsymbol{\theta}_{i}}\left(\boldsymbol{a}\mid\dot{\boldsymbol{s}}\right)\right],\forall k,i\text{.}
\end{equation}
Then, the asymptotic consistency of gradients can be decomposed into
the following thee steps:
\begin{align}
\mathrm{Step}1: & \underset{t\rightarrow\infty}{\mathrm{lim}}\bigl\Vert\nabla_{\boldsymbol{\theta}}\hat{f}_{k}\left(\boldsymbol{\theta}_{i}\right)-\nabla_{\boldsymbol{\theta}}\dot{f}_{k}\left(\boldsymbol{\theta}_{i}\right)\bigr\Vert_{2}=0,\label{eq:lim_g_step1}\\
\mathrm{Step}2: & \underset{t\rightarrow\infty}{\mathrm{lim}}\bigl\Vert\hat{\bar{\boldsymbol{g}}}_{k,i}-\nabla_{\boldsymbol{\theta}}\hat{f}_{k}\left(\boldsymbol{\theta}_{i}\right)\bigr\Vert_{2}=0,\label{eq:lim_g_step2}\\
\mathrm{Step}3: & \underset{t\rightarrow\infty}{\mathrm{lim}}\bigl\Vert\hat{\boldsymbol{g}}_{k,i}-\hat{\bar{\boldsymbol{g}}}_{k,i}\bigr\Vert_{2}=\epsilon_{Q}.\label{eq:lim_g_step3}
\end{align}
According to the definations of $\nabla_{\boldsymbol{\theta}}\hat{f}_{k}\left(\boldsymbol{\theta}_{i}\right)$,
$\nabla_{\boldsymbol{\theta}}f_{k}\left(\boldsymbol{\theta}_{i}\right)$,
$\hat{\boldsymbol{g}}_{k,i}$, and $\hat{\bar{\boldsymbol{g}}}_{k,i}$,
it's easy to obtain that 
\begin{align*}
\bigl\Vert\nabla_{\boldsymbol{\theta}}\hat{f}_{k}\left(\boldsymbol{\theta}_{i}\right)-\nabla_{\boldsymbol{\theta}}\dot{f}_{k}\left(\boldsymbol{\theta}_{i}\right)\bigr\Vert_{2}= & O\bigl(\bigl\Vert\hat{f}_{k,i}-\dot{f}_{k}\left(\boldsymbol{\theta}_{i}\right)\bigr\Vert_{2}\bigr),
\end{align*}
\begin{align*}
\bigl\Vert\hat{\boldsymbol{g}}_{k,i}-\hat{\bar{\boldsymbol{g}}}_{k,i}\bigr\Vert_{2}= & O\bigl(\bigl\Vert Q_{\boldsymbol{\omega}_{k,i}}\left(\dot{\boldsymbol{s}},\boldsymbol{a}\right)-\hat{Q}_{k}^{\pi_{\boldsymbol{\theta}_{i}}}\left(\dot{\boldsymbol{s}},\boldsymbol{a}\right)\bigr\Vert_{2}\bigr)=\epsilon_{Q}.
\end{align*}
Therefore, Step 1 holds as long as the asymptotic consistency of the
function values holds, and Step 3 clearly holds.

In the following, we focus on proving the Step 2 in (\ref{eq:lim_g_step2}).
This proof relies on a technical lemma \cite[Lemma 1]{Lemma3}, which
is restated below for completeness. \begin{lemma} Let $(\Omega,\mathcal{G},\mathbb{P})$
denote a probability space and let $\{\mathcal{G}_{t}\}$ denote an
increasing sequence of $\sigma$-field contained in $\mathcal{G}$.
Let $\{z^{t}\}$, $\{w^{t}\}$ be sequences of $\mathcal{G}_{t}$-measurable
random vectors satisfying the relations\vspace{-0.2cm}
\begin{align}
w^{t+1}= & \prod_{\mathcal{W}}(w^{t}+\alpha_{t}(\varrho^{t}-w^{t}))\\
\mathbb{E}[\varrho^{t}|\mathcal{G}_{t}]= & z^{t}+o^{t}\nonumber 
\end{align}
where $\alpha_{t}\geq0$ and the set $\mathcal{W}$ is convex and
closed, $\prod_{\mathcal{W}}(\cdot)$ denotes projection on $\mathcal{W}$.
Let\\
(a) all accumulation points of $\{w^{t}\}$ belong to $\mathcal{W}$
w.p.l.,\\
(b) there is a constant $C$ such that $\mathbb{E}[\Vert\varrho^{t}\Vert_{2}|\mathcal{G}_{t}]\leq C$,
$\forall t\geq0$,\\
(c) $\sum_{t=0}^{\infty}\mathbb{E}[(\alpha_{t})^{2}+\alpha_{t}\Vert o^{t}\Vert]\text{<}\infty$\\
(d) $\sum_{t=0}^{\infty}\alpha_{t}=\infty$, and (e) $\Vert z^{t+1}-z^{t}\Vert/\alpha_{t}\rightarrow0$
w.p.l.,\\
Then $z^{t}-w^{t}\rightarrow0\ w.p.1.$\end{lemma}

Since the step size $\left\{ \alpha_{t}\right\} $ follows Assumption
2, and $\dot{C}_{i},\forall i$ is bounded, it's easy to prove that
the conditions (a), (b), and (d) in the Lemma above are satisfied.
Now, we are ready to prove the technical condition (c).

Recalling the defination of $\tilde{\bar{\boldsymbol{g}}}_{k,i}$
and $\nabla_{\boldsymbol{\theta}}\hat{f}_{k}\left(\boldsymbol{\theta}_{i}\right)$,
we obtain the stochastic policy gradient error:
\begin{align}
 & \left\Vert o^{t}\right\Vert _{2}=\bigl\Vert\tilde{\bar{\boldsymbol{g}}}_{k,i}-\nabla_{\boldsymbol{\theta}}\hat{f}_{k}\left(\boldsymbol{\theta}_{i}\right)\bigr\Vert_{2}\label{eq:stochastic bias}\\
\overset{a}{=} & \frac{1}{B}\stackrel[j=Bi'+1]{Bi'+B}{\sum}O\bigl(\bigl\Vert P_{S}\bigl(\dot{\boldsymbol{s}}_{j}\bigr)-\mathbf{P}_{\pi_{\boldsymbol{\theta}_{i'+1}}}\bigl(\dot{\boldsymbol{s}}_{j}\bigr)\bigr\Vert_{\mathrm{TV}}.\nonumber 
\end{align}
According to (32)-(34) in Appendix A of our manuscript and Assumption
2, it's easy to prove that condition (c) is held. Moreover, for the
condition (e), we have
\begin{align}
 & \bigl\Vert\nabla_{\boldsymbol{\theta}}\hat{f}_{i}\left(\boldsymbol{\theta}_{t+1}\right)-\nabla_{\boldsymbol{\theta}}\hat{f}_{i}\left(\boldsymbol{\theta}_{t}\right)\bigr\Vert_{2}\\
= & O\bigl(\bigl\Vert\boldsymbol{\theta}_{t+1}-\boldsymbol{\theta}_{t}\bigr\Vert_{2}+\bigl\Vert\mathbf{P}_{\pi_{\boldsymbol{\theta}_{t+1}}}-\mathbf{P}_{\pi_{\boldsymbol{\theta}_{t}}}\bigr\Vert_{TV}\bigr)=O\bigl(\beta_{t}\bigr)\nonumber 
\end{align}
It can be seen from Assumption 2 that technical condition (e) is also
satisfied. This completes the proof of Step 2.

\end{appendices}

\bibliographystyle{IEEEtran}
\bibliography{RLreferences}

\begin{thebibliography}{10}
\providecommand{\url}[1]{#1}
\csname url@samestyle\endcsname
\providecommand{\newblock}{\relax}
\providecommand{\bibinfo}[2]{#2}
\providecommand{\BIBentrySTDinterwordspacing}{\spaceskip=0pt\relax}
\providecommand{\BIBentryALTinterwordstretchfactor}{4}
\providecommand{\BIBentryALTinterwordspacing}{\spaceskip=\fontdimen2\font plus
\BIBentryALTinterwordstretchfactor\fontdimen3\font minus
  \fontdimen4\font\relax}
\providecommand{\BIBforeignlanguage}[2]{{%
\expandafter\ifx\csname l@#1\endcsname\relax
\typeout{** WARNING: IEEEtran.bst: No hyphenation pattern has been}%
\typeout{** loaded for the language `#1'. Using the pattern for}%
\typeout{** the default language instead.}%
\else
\language=\csname l@#1\endcsname
\fi
#2}}
\providecommand{\BIBdecl}{\relax}
\BIBdecl

\bibitem{XR0large0packet}
F.~Alriksson, D.~H. Kang, C.~Phillips, J.~L. Pradas, and A.~Zaidi, ``Xr and 5g:
  Extended reality at scale with time-critical communication,'' \emph{Ericsson
  Technology Review}, vol. 2021, no.~8, pp. 2--13, 2021.

\bibitem{XR}
X.~Zhao, Y.-J.~A. Zhang, M.~Wang, X.~Chen, and Y.~Li, ``Online multi-user
  scheduling for xr transmissions with hard-latency constraint: Performance
  analysis and practical design,'' \emph{IEEE Trans. Commun.}, 2024.

\bibitem{resource0consumption}
B.~Chang, L.~Zhang, L.~Li, G.~Zhao, and Z.~Chen, ``Optimizing resource
  allocation in urllc for real-time wireless control systems,'' \emph{IEEE
  Trans. Veh. Technol.}, vol.~68, no.~9, pp. 8916--8927, 2019.

\bibitem{EEbook}
Z.~Wei, Y.~Cai, D.~W.~K. Ng, and J.~Yuan, ``Energy-efficient radio resource
  management,'' \emph{Wiley 5G Ref: The Essential 5G Reference Online}, pp.
  1--23, 2019.

\bibitem{EE}
R.~Ruby, S.~Zhong, D.~W.~K. Ng, K.~Wu, and V.~C. Leung, ``Enhanced
  energy-efficient downlink resource allocation in green non-orthogonal
  multiple access systems,'' \emph{Computer Communications}, vol. 139, pp.
  78--90, 2019.

\bibitem{SCAOPO}
C.~Tian, A.~Liu, G.~Huang, and W.~Luo, ``Successive convex approximation based
  off-policy optimization for constrained reinforcement learning,'' \emph{IEEE
  Trans. Signal Process.}, vol.~70, pp. 1609--1624, 2022.

\bibitem{average-delay1}
R.~A. Berry and R.~G. Gallager, ``Communication over fading channels with delay
  constraints,'' \emph{IEEE Trans. Inf. Theory}, vol.~48, no.~5, pp.
  1135--1149, 2002.

\bibitem{average-delay2}
Y.~Mao, J.~Zhang, S.~H. Song, and K.~B. Letaief, ``Stochastic joint radio and
  computational resource management for multi-user mobile-edge computing
  systems,'' \emph{IEEE Trans. Wireless Commun.}, vol.~16, no.~9, pp.
  5994--6009, 2017.

\bibitem{heuristic1-8}
S.~Shakkottai and R.~Srikant, ``Scheduling real-time traffic with deadlines
  over a wireless channel,'' in \emph{Proc. 2nd ACM international workshop on
  Wireless mobile multimedia}, 1999, pp. 35--42.

\bibitem{non-causal}
W.~Chen, M.~J. Neely, and U.~Mitra, ``Energy efficient scheduling with
  individual packet delay constraints: Offline and online results,'' in
  \emph{Proc IEEE Int. Conf. Comput. Commun.}, 2007, pp. 1136--1144.

\bibitem{AWGN}
M.~Khojastepour and A.~Sabharwal, ``Delay-constrained scheduling: power
  efficiency, filter design, and bounds,'' in \emph{Proc IEEE Int. Conf.
  Comput. Commun.}, vol.~3, 2004, pp. 1938--1949 vol.3.

\bibitem{Poisson}
I.~Fawaz, M.~Sarkiss, and P.~Ciblat, ``Optimal resource scheduling for energy
  harvesting communications under strict delay constraint,'' in \emph{Proc IEEE
  Int. Conf. Commun.}, 2018, pp. 1--6.

\bibitem{model-free-CPO}
M.~Sulaiman, M.~Ahmadi, M.~A. Salahuddin, R.~Boutaba, and A.~Saleh,
  ``Generalizable resource scaling of 5{G} slices using constrained
  reinforcement learning,'' in \emph{Proc. NOMS IEEE/IFIP Netw. Operations
  Manage. Symp.}\hskip 1em plus 0.5em minus 0.4em\relax IEEE, 2023, pp. 1--9.

\bibitem{DRL}
K.~Arulkumaran, M.~P. Deisenroth, M.~Brundage, and A.~A. Bharath, ``Deep
  reinforcement learning: A brief survey,'' \emph{IEEE Trans. Signal Process.},
  vol.~34, no.~6, pp. 26--38, 2017.

\bibitem{PPOLagTRPOLag}
J.~Achiam and D.~Amodei, ``Benchmarking safe exploration in deep reinforcement
  learning,'' 2019.

\bibitem{NSGD1}
S.~Khodadadian, T.~T. Doan, J.~Romberg, and S.~T. Maguluri, ``Finite sample
  analysis of two-time-scale natural actor-critic algorithm,'' \emph{IEEE
  Trans. Automat. Contr.}, 2022.

\bibitem{primal_dual_conv2}
D.~Ding, K.~Zhang, T.~Basar, and M.~Jovanovic, ``Natural policy gradient
  primal-dual method for constrained markov decision processes,''
  \emph{Advances in Neural Information Processing Systems}, vol.~33, pp.
  8378--8390, 2020.

\bibitem{ding2024last}
D.~Ding, C.-Y. Wei, K.~Zhang, and A.~Ribeiro, ``Last-iterate convergent policy
  gradient primal-dual methods for constrained mdps,'' \emph{Advances in Neural
  Information Processing Systems}, vol.~36, 2024.

\bibitem{SCAOPO18}
D.~Ding, X.~Wei, Z.~Yang, Z.~Wang, and M.~Jovanovic, ``Provably efficient safe
  exploration via primal-dual policy optimization,'' in \emph{International
  Conference on Artificial Intelligence and Statistics}.\hskip 1em plus 0.5em
  minus 0.4em\relax PMLR, 2021, pp. 3304--3312.

\bibitem{CPO}
J.~Achiam, D.~Held, A.~Tamar, and P.~Abbeel, ``Constrained policy
  optimization,'' in \emph{ICML}.\hskip 1em plus 0.5em minus 0.4em\relax PMLR,
  2017, pp. 22--31.

\bibitem{ProjectionCPO}
T.-Y. Yang, J.~Rosca, K.~Narasimhan, and P.~J. Ramadge, ``Projection-based
  constrained policy optimization,'' \emph{arXiv preprint arXiv:2010.03152},
  2020.

\bibitem{firstorderCPO}
Y.~Zhang, Q.~Vuong, and K.~Ross, ``First order constrained optimization in
  policy space,'' \emph{Adv. Neural Inf. Process. Syst.}, vol.~33, pp.
  15\,338--15\,349, 2020.

\bibitem{reviewer1_1_2}
S.~Paternain, L.~Chamon, M.~Calvo-Fullana, and A.~Ribeiro, ``Constrained
  reinforcement learning has zero duality gap,'' in \emph{Advances in Neural
  Information Processing Systems}, H.~Wallach, H.~Larochelle, A.~Beygelzimer,
  F.~d\textquotesingle Alch\'{e}-Buc, E.~Fox, and R.~Garnett, Eds.,
  vol.~32.\hskip 1em plus 0.5em minus 0.4em\relax Curran Associates, Inc.,
  2019.

\bibitem{meta0learning1}
H.~Hu, G.~Huang, X.~Li, and S.~Song, ``Meta-reinforcement learning with dynamic
  adaptiveness distillation,'' \emph{IEEE Trans. Neural Netw. Learn. Syst.},
  vol.~34, no.~3, pp. 1454--1464, 2021.

\bibitem{nonstationaryURLLC}
W.~Wu, J.~Dong, Y.~Sun, and F.~R. Yu, ``Heterogeneous markov decision process
  model for joint resource allocation and task scheduling in network slicing
  enabled internet of vehicles,'' \emph{IEEE Wireless Communications Letters},
  vol.~11, no.~6, pp. 1118--1122, 2022.

\bibitem{MAML}
C.~Finn, P.~Abbeel, and S.~Levine, ``Model-agnostic meta-learning for fast
  adaptation of deep networks,'' in \emph{International conference on machine
  learning}.\hskip 1em plus 0.5em minus 0.4em\relax PMLR, 2017, pp. 1126--1135.

\bibitem{MAML-CPO}
M.~Cho and C.~Sun, ``Constrained meta-reinforcement learning for adaptable
  safety guarantee with differentiable convex programming,'' in
  \emph{Proceedings of the AAAI Conference on Artificial Intelligence},
  vol.~38, no.~19, 2024, pp. 20\,975--20\,983.

\bibitem{gradientbased}
A.~Nichol, J.~Achiam, and J.~Schulman, ``On first-order meta-learning
  algorithms,'' \emph{arXiv preprint arXiv:1803.02999}, 2018.

\bibitem{PERAL}
K.~Rakelly, A.~Zhou, C.~Finn, S.~Levine, and D.~Quillen, ``Efficient off-policy
  meta-reinforcement learning via probabilistic context variables,'' in
  \emph{International conference on machine learning}.\hskip 1em plus 0.5em
  minus 0.4em\relax PMLR, 2019, pp. 5331--5340.

\bibitem{context3}
P.~Jiang, S.~Song, and G.~Huang, ``Exploration with task information for meta
  reinforcement learning,'' \emph{IEEE Trans. Neural Netw. Learn. Syst.},
  vol.~34, no.~8, pp. 4033--4046, 2021.

\bibitem{VAE}
A.~Xie, J.~Harrison, and C.~Finn, ``Deep reinforcement learning amidst
  continual structured non-stationarity,'' in \emph{International Conference on
  Machine Learning}.\hskip 1em plus 0.5em minus 0.4em\relax PMLR, 2021, pp.
  11\,393--11\,403.

\bibitem{RZF}
R.~Zakhour and S.~V. Hanly, ``Base station cooperation on the downlink: Large
  system analysis,'' \emph{IEEE Trans. Inf. Theory}, vol.~58, no.~4, pp.
  2079--2106, 2012.

\bibitem{Gaussianpolicy}
R.~S. Sutton and A.~G. Barto, \emph{Reinforcement learning: An
  introduction}.\hskip 1em plus 0.5em minus 0.4em\relax MIT press, 2018.

\bibitem{DQlearning}
P.~Xu and Q.~Gu, ``A finite-time analysis of {Q}-learning with neural network
  function approximation,'' in \emph{ICML}, 2020, pp. 10\,555--10\,565.

\bibitem{modelbased}
Q.~Huang, ``Model-based or model-free, a review of approaches in reinforcement
  learning,'' in \emph{2020 International Conference on Computing and Data
  Science (CDS)}, 2020, pp. 219--221.

\bibitem{meanfield}
W.~Han and Y.~Yang, ``Statistical inference in mean-field variational bayes,''
  \emph{arXiv preprint arXiv:1911.01525}, 2019.

\bibitem{reward0shaping1}
H.~Zou, T.~Ren, D.~Yan, H.~Su, and J.~Zhu, ``Reward shaping via
  meta-learning,'' \emph{arXiv preprint arXiv:1901.09330}, 2019.

\bibitem{linnerAC}
S.~Qiu, Z.~Yang, J.~Ye, and Z.~Wang, ``On finite-time convergence of
  actor-critic algorithm,'' \emph{IEEE J. Sel. Areas Inf. Theory}, vol.~2,
  no.~2, pp. 652--664, 2021.

\bibitem{Yangrui}
A.~Liu, R.~Yang, T.~Q.~S. Quek, and M.-J. Zhao, ``Two-stage stochastic
  optimization via primal-dual decomposition and deep unrolling,'' \emph{IEEE
  Trans. Signal Process.}, vol.~69, pp. 3000--3015, 2021.

\bibitem{harddelay4}
I.~Fawaz, M.~Sarkiss, and P.~Ciblat, ``Optimal resource scheduling for energy
  harvesting communications under strict delay constraint,'' in \emph{2018 IEEE
  International Conference on Communications (ICC)}, 2018, pp. 1--6.

\bibitem{model0free0RL1}
A.~T.~Z. Kasgari and W.~Saad, ``Model-free ultra reliable low latency
  communication (urllc): A deep reinforcement learning framework,'' in
  \emph{ICC 2019-2019 IEEE International Conference on Communications
  (ICC)}.\hskip 1em plus 0.5em minus 0.4em\relax IEEE, 2019, pp. 1--6.

\bibitem{Lemma3}
A.~Ruszczy{\'n}ski, ``Feasible direction methods for stochastic programming
  problems,'' \emph{Mathematical Programming}, vol.~19, no.~1, pp. 220--229,
  1980.

\end{thebibliography}

\end{document}